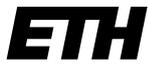

Eidgenössische Technische Hochschule Zürich
Swiss Federal Institute of Technology Zurich

Institut für Umweltingenieurwissenschaften
Ökologisches Systemdesign

ETH Zurich
John-von-Neumann-Weg 9
8093 Zürich
Switzerland

Phone +41 44 633 44 78
https://esd.ifu.ethz.ch/

# Multiscale Orientation Values for Biodiversity, Climate and Water: A Scientific Input for Science-Based Targets

Multiscale Orientation Values for Biodiversity, Climate and Water: A Scientific Input for Science-Based Targets

# Additional Information

## Individuals and Departments Involved

| Name | Department / Responsibility |
|---|---|
| **Authors** | |
| Dr Viktoras Kulionis | ESD, ETH Zürich |
| Dr Stephan Pfister | ESD, ETH Zürich |
| Dr Andreas Froemelt | ESD, ETH Zürich |
| **Advisory group** | |
| Andreas Hauser | FOEN, Economics and Innovation Division |
| Niklas Nierhoff | FOEN, Economics and Innovation Division |
| Dr Glenn Litsios | FOEN, Biodiversity and Landscape Division |
| Dr Andreas Bachmann | FOEN, Soil and Biotechnology Division |
| Dr Roger Ramer | FOEN, Climate Division |

# Table of Contents

# Index of Tables and Figures

## Tables

## Figures

Multiscale Orientation Values for Biodiversity, Climate and Water: A Scientific Input for Science-Based Targets

Multiscale Orientation Values for Biodiversity, Climate and Water: A Scientific Input for Science-Based Targets

# 1. Introduction

## 1.1. Background

Over the past three centuries, human-induced effects on the global environment have escalated (Crutzen, 2002). Experts suggest that the Earth System has been modified and may have entered a new human-dominated geological epoch, the Anthropocene (Lewis and Maslin, 2015). Further pressure on the Earth System could de-stabilise critical planetary processes and trigger abrupt or irreversible environmental changes that would be harmful or even catastrophic for human wellbeing (Rockström et al., 2009). There is a need for concerted global efforts to reduce environmental impacts and guarantee a safe operating space needed to support global sustainable development and wellbeing of humankind.

The Planetary Boundaries (PB) framework was proposed as a guidance system for maintaining a state of the Earth System that resembles the Holocene epoch. The framework defines boundaries for nine fundamental global processes that together regulate the stability of the Earth system. In this respect, the PBs jointly define a global "safe operating space" in which human society can develop and thrive. For each of the PBs, so-called "control variables" proxies to measure whether they are transgressed on the global level have been defined (Rockström et al., 2009; Steffen et al., 2015). Four of the nine PBs are currently transgressed due to human activity: Climate change, biosphere integrity, land system change and biogeochemical flows. Climate change and biosphere integrity are recognised as "core" PBs based on their fundamental importance for the Earth System.

Since its introduction in 2009, the PB framework has generated considerable interest within the policy, governance, and business sectors to inform efforts toward global sustainability (Steffen et al., 2015). It has provided input to the United Nations' seventeen Sustainable Development Goals (SDGs) and has informed EU policy (EEA-FOEN, 2020) and national policies and targets in several countries. The Swedish Environmental Protection Agency commissioned the first study to examine whether and how the planetary boundaries concept could inform Swedish environmental policy objectives (Nykvist et al., 2013). Germany's 'Integrated Environmental Programme 2030' (BMUB, 2016) identifies the need to operate within planetary boundaries as a key priority. In Switzerland, the planetary boundary concept is explicitly part of the 2016-2019 sustainable development strategy (Swiss Federal Council, 2016), and environmental footprints of Switzerland are regularly monitored against planetary boundaries (Dao et al., 2015, Frischknecht et al., 2018). The Netherlands Environment Assessment Agency (PBL) support the national implementation of environment-related SDGs using the concept of planetary boundaries (Lucas and Wilting, 2018). On a city level, Amsterdam, among other cities, embraced the so-called Doughnut model (see, e.g., Raworth, 2020), combining planetary with social boundaries.

The Planetary Boundaries are global in scale and are not designed to be "downscaled" or "disaggregated" to smaller levels, such as regional or local ecosystems. However, the decisions about resource use and environmental policies are usually made at a sub-global scale (e.g., countries, basins, and regions). Therefore, there is a growing need for the Planetary Boundaries to be translated from their global-scale viewpoint to support sustainability decision-making at other levels.

Various equity principles such as responsibility, capability, equality, sovereignty (Höhne et al., 2014) have been applied as the basis for effort-sharing approaches and distribution of global Planetary Boundaries to specific national or regional contexts. Although there is little consistency in the procedures these studies use, common elements of a universal PB operationalisation method are beginning to emerge (see, e.g., Dao et al., 2015; Häyhä et al., 2016; Lucas and Wiliting, 2020; EEA-FOEN, 2020). Building on the previous studies here, we explore a range of aspects relevant for the operationalization of the Planetary Boundaries framework.

## 1.2. Aims

This report aims to evaluate and further develop orientation values for the biodiversity, climate change and freshwater use planetary boundaries on the scales of countries, economic sectors, and cities. Our understanding of the term "orientation values" within this paper is that these values provide a scientific orientation which is useful as a starting point for setting targets on multiple scales such as countries, cities and economic sectors.

This study aims to:

- Provide an overview of existing attempts to derive orientation values for the climate, biodiversity and water planetary boundaries

- Provide an overview of existing allocation approaches

- Based on the previous literature, propose a methodology to allocate planetary boundaries to different sub-global scales and sectors

- Present a global and country-specific status quo analysis for the selected environmental indicators
    - climate change,
    - freshwater use
    - land-use related biodiversity loss,

- Apply the proposed methodology to derive country, city and sector-specific budgets consistent with the planetary boundaries concept

# 2. Methodology

Translating PBs into sub-global levels is highly normative and involves multiple considerations. Häyä et al. (2016) proposed a framework for translating the PBs into national-level "fair shares" of Earth's safe operating space, emphasising the need to take into account the biophysical, socio-economic, and ethical dimensions:

- The biophysical dimension deals with the geographical scales of the PBs processes and their interactions.

- The socio-economic dimension addresses the sub-global links created by production and consumption patterns and through international trade.

- The ethical dimension addresses equity in sharing the global safe operating space and recognising the differences between countries' rights, abilities, and responsibilities.

The procedure applied in this report to assess the environmental performance of countries, cities and industries involves three steps:

1) Derive safe operating space (SOS) for a specific country (in this case Switzerland) and a particular indicator;

2) Identify the current environmental state (e.g., current Greenhouse gas emissions);

3) Evaluate if a country/city or industry operates within an allocated SOS for a given environmental indicator. These steps are explained in more detail below and shown in Figure 1.

**Multiscale Orientation Values for Biodiversity, Climate and Water: A Scientific Input for Science-Based Targets**

Figure 1 The Planetary Boundaries allocation framework

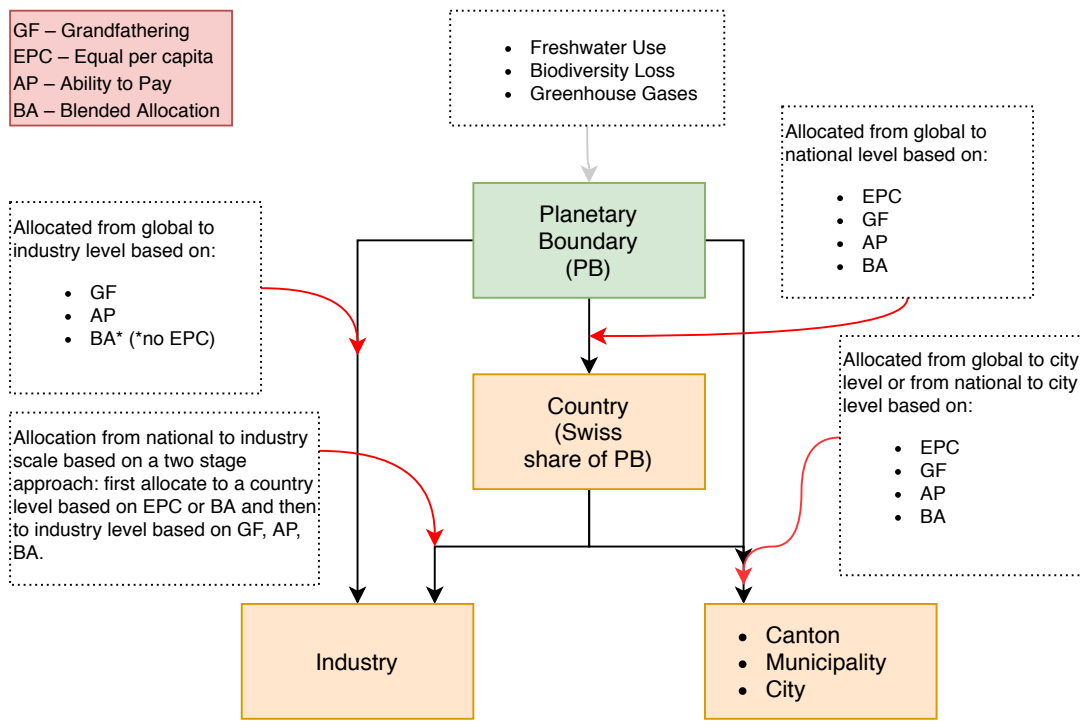

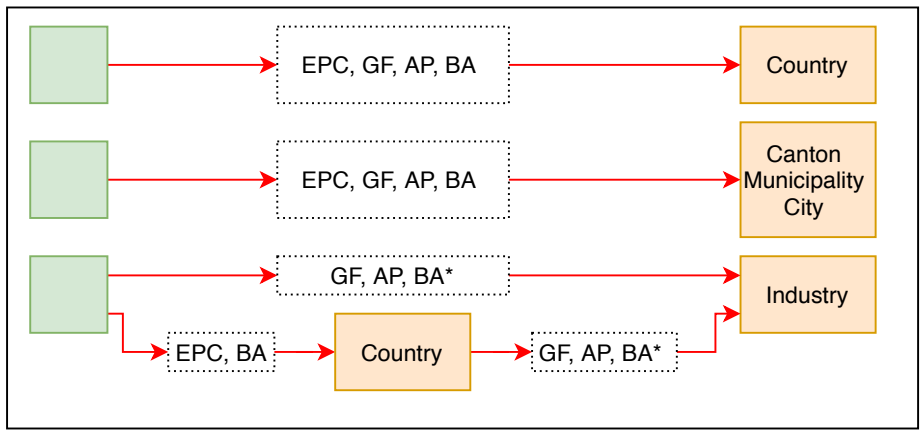

## 2.1. Biophysical dimension

In this report, we assess three PBs: climate change, biosphere integrity and freshwater use. Climate change PB can be characterised as truly global phenomena because it is the total amount of greenhouse gas (GHG) emissions that are important, not the location of the emissions. On the other hand, biodiversity integrity and freshwater use are local or regional phenomena that can accumulate to a global level. This implies that the latter PBs cannot only be directly connected to a single, well-mixed universal indicator and is better assessed with the use of multiple regional indicators.

Steffen et al. (2015) propose complementing the global limits with sub-global limits for five planetary boundaries: functional diversity (as part of biosphere integrity), phosphorus (as part of biogeochemical flows), land system change, freshwater use and atmospheric aerosol loading.

For the climate change PB, we use global boundary, while for the freshwater use and biosphere integrity PBs we use local/regional and global level boundaries. For water, this is achieved by taking into account water use at the watershed level (a similar approach was applied in Bjørn et al., 2020). These watersheds are large watersheds with an outflow to the sea (or salt lake), such as the Rhine and can be considered global watersheds.

It is important to note that PB processes, control variables and boundary levels also have different characteristics. Some are defined in terms of global stocks or budgets, others as flows over time some are expressed in terms of changes in the biophysical state, others in terms of the anthropogenic driver (Dao et al., 2015).

### 2.1.1. Climate change

The climate change planetary boundary aims at minimizing the risk of non-linear, possibly abrupt and irreversible tipping points in the Earth System – such as the loss of the Amazon rainforest or the West Antarctic ice sheet. The proposed control variables in the PB framework are atmospheric $CO_2$ concentrations and radiative forcing (Steffen et al., 2015).

Global political targets are expressed in terms of maximum allowable temperature increase. The Paris Agreement set an international political goal to keep the rise of global mean temperature by 2100 to less than 2°C, and closer to 1.5°C, relative to 1861–80 temperatures. A global carbon budget defines a maximum amount of $CO_2$ emissions that could globally be emitted this century, while still have a likely chance of limiting global temperature rise to 1.5°–2°C above pre-industrial levels. From 2020 onwards this translates to a remaining cumulative carbon budget of 1315 Gt $CO_2$ for limiting warming to 2.0°C with a 50% probability (or 985 Gt $CO_2$ for a 66% probability). To remain within a 1.5°C (with a 50% probability) degree pathway, the remaining carbon budget is reduced to 395 Gt $CO_2$ (235 Gt $CO_2$ for a 66% probability). The carbon budget is defined for carbon emissions, and including other greenhouse gases (methane and nitrous oxide) increases global carbon budget by roughly 25% (Willet et al., 2019).

### 2.1.2. Freshwater use

The freshwater use boundary aims to provide a limit to a global water cycle modification. Most previous attempts to translate the water PB to sub-global scales has primarily adopted either a "fair shares" approach (global boundary) or a local safe operating space approach (local level boundary, e.g., watershed).

Both of these approaches have their strengths and weaknesses. The local safe operating space approach quantifies local limits to water system modifications (mainly environmental flow requirements) but does not provide information about potential impacts beyond the local context being considered. Notice that, both global and local/regional boundaries can be allocated to consumer/producers depending on what question one seeks to answer.

The "fair shares" approach complements the local safe operating space approach by providing a tool for systematic comparisons among regions or countries, assessing global responsibility, and allocating responsibility for local contribution to global processes. However, the fair shares approach does not provide guidance for whether the water cycle remains within locally relevant limits, which is the primary concern of water managers and policymakers. Therefore, it requires integration with the local safe operating space approach. Zipper et al. (2020) desribed a cross-scale approach that combines the advantages of the fair shares approach (Earth System relevance and global responsibility) and the local safe operating space approach (local relevance). This approach aims to ensure that actions in a local context are consistent with both local and global water sustainability.

It should be noted that the freshwater use boundary itself has been critiqued as a singular measure that does not adequately represent all types of human interference with the complex global water cycle and Earth System. Recently, Gleeson et al. (2020) proposed to define a new water planetary boundary consisting of six water sub-boundaries (atmospheric water, frozen water, groundwater, soil moisture, and surface water) relating to the major stores of freshwater to more holistically represent the various functions of water in maintaining Earth System stability. The original freshwater use boundary is retained in this approach as the surface water sub-boundary, together with new water planetary sub-boundaries for frozen water, groundwater, soil moisture, and two sub-boundaries for different aspects of atmospheric water. Gleeson et al. (2020) suggested potential control and response variables for the new sub-boundaries, but significant work remains to select and evaluate appropriate variables and boundary values. One major issue is the limited detail available in global hydrological models.

In this report, we do not investigate the new water sub-boundaries introduced by Glesson et al. (2020) and focus on the original freshwater use boundary by Steffen et al. (2015), for which data is available. However, we do take into account local/regional characteristics into consideration as suggested by Zipper et al. (2020), i.e., the fair shares approach (Earth System relevance and global responsibility) and the local safe operating space approach (local/regional relevance).

### 2.1.3. Biodiversity loss

Species extinction is a natural process and would occur without human intervention, however, human actions have dramatically accelerated biodiversity loss. In the PB framework, biodiversity loss is considered one of two "key" PBs that plays a crucial rule in the Earth System functioning. For instance, loss of species can increase the vulnerability of terrestrial and aquatic ecosystems to changes in climate and ocean acidity, thus reducing the safe boundary levels for these processes (Rökstrom et al., 2009).

**Multiscale Orientation Values for Biodiversity, Climate and Water: A Scientific Input for Science-Based Targets**

Biodiversity is usually considered a regional issue rather than a global issue since changes occur at a local or regional scale. A global perspective can, however, be adopted since evidence for the important role of biodiversity for ecosystem functioning and human wellbeing is considerable (Hooper et al., 2005; Cardinale et al., 2012) and through migration and dispersal of species, ecosystems might benefit from biodiversity in other regions. The global PB may aim at preventing global extinction of threatened species, and additional regional/local boundaries would warrant that less valuable ecosystems (with less globally threatened or endemic species) are preserved and remain functioning.

Steffen et al. (2015) proposed a two-component approach to account for regional heterogeneity and to better capture key roles of the biosphere in the Earth system. The biodiversity intactness index and global species extinction rate were proposed as the control variables. Both measures are considered to be transgressed globally, and increased efforts are required to prevent further losses. However, both variables are regarded as an interim solution, and the scientific community is actively seeking more appropriate control variables.

Various indicators have been proposed and used in the attempts to allocate the biodiversity PB. Lucas et al. (2020) applied mean species abundance (MSA) as an indicator of functional diversity to assess environmental performance with respect to the selected PBs of for the EU, US, China and India. The MSA measures the mean abundance of species in a disturbed situation relative to their mean abundance in an undisturbed reference situation. Frischknecht et al. (2018) applied potentially disappearing fraction (PDF) of global species years as an indicator to capture biodiversity loss, following LCA impact assessment methods (Chaudhary et al. 2015, 2016). An extinction rate of 10 species per million species and per year over the last 1500 years (1.5% species lost over 1500 years), was assumed as the threshold value, based on Steffen et al. (2015).

Rounsevell et al. (2020) proposed a similar approach based on mean species abundance (MSA) as an indicator and 20 species extinctions per million species as a target. The target is motivated by the fact that over the coming decades, some continuing loss of species is inevitable given the current human domination of Earth's systems, so the authors suggest an ambitious but achievable rate: keeping described species extinctions to well below 20 per year over the next 100 years. Thereafter, a rate closer to background rates (i.e., prehistorical rates) should be the aspiration. Marquardt et al. (2019) examined differences between alpha (local within site diversity, e.g., MSA) and gamma (global across site diversity, e.g., global PDF) biodiversity indicators. They found that different alpha indicators show close alignment, but there was limited convergence between alpha and gamma biodiversity footprints. Their results highlight the relevance of including both alpha (MSA) and gamma (PDF) diversity indicators in biodiversity footprint calculations.

Although biodiversity loss from human activities is widely known, well-defined and measurable targets (analogous to the Paris agreement) have not been agreed yet on an international level. While climate scientists arrived at a single numerical target for maintaining Earth's atmosphere at safe limits, biodiversity scientists identified multiple targets to prevent species extinctions and the rapid erosion of biodiversity. However, it is argued that for effective communication, biodiversity needs should be encompassed within a single target (Dinerstein et al., 2019). Therefore, the aspirational goal of 50% protected area has emerged under the name Nature Needs Half (NNH) (Locke, 2014; Dinerstein et al., 2017). Protected areas are considered the cornerstone for habitat and species conservation. Studies document that well-managed reserves are far more effective in safeguarding biodiversity than other forms of land use (Gray, 2016). NNH is based on four fundamental goals: 1) represent all native ecosystem types and successional stages across their natural range of variation, (2) maintain viable populations of all native species in natural patterns of abundance and distribution, (3) maintain ecological function and ecosystem services, and (4) address environmental change to maintain evolutionary processes and adapt to the impacts of climate change.

Dinerstein et al. (2017) assessed progress towards the protection of 50% of the terrestrial biosphere within the ecoregion framework. They evaluated protection statuses of ecoregions of the world and measured progress towards half protected (50% or more of the total ecoregion area) goal and the remaining natural habitat.

We use their results as a proxy to derive a safe operating space (SOS) in each ecoregion and evaluate Swiss land use and biodiversity loss in each ecoregion relative to SOS. This allows identifying potential ecoregion-hotspots where improvements are needed the most.

Just like many other approaches, NNH has strengths and weaknesses. One of the key strengths of this approach is that it offers a very ambitious yet straightforward and easy to communicate goal. A potential pitfall is that larger protection areas may not necessarily translate into declining biodiversity loss and increasing species richness (Visconti et al., 2019). Rounsevell et al. (2020) emphasised that many comprehensive proposals for the post-2020 agenda focus on achieving conservation actions, such as increasing the coverage of areas dedicated to wildlife, or maintaining intact wilderness, rather than specifying required outcomes for biodiversity.

## 2.2. Socio-economic dimension (accounting principles)

Production-based accounting (PBA) and consumption-based accounting (CBA) are two of the most commonly used methods to measure environmental pressures/impacts. The PBA method is the primary method used by statistical offices and international organisations; for instance, it is currently the UNFCCC's adopted accounting principle.

However, international fragmentation of production processes has raised the awareness for the need to complement the PBA approach, with other accounting approaches (separation of production and consumption activities). The CBA, with the calculation of country environmental footprints, has emerged as one of the options. The difference between the two accounting approaches centres on how they account for international trade. Both methods have their strengths and weaknesses, and which one is better depends on the question at hand (see, e.g., Afionis et al., 2017). Note that throughout this text, we use terms consumption-based accounting, CBA and footprint interchangeably.

The PBA accounts for all environmental pressures that result from the economic activities (i.e., production of good and services) of a country's resident companies and private households (together known as "resident" units) irrespective of the geographic location where their activities take place (EEA, 2013). This approach accounts for environmental impacts related to exports (green box in Figure 2). It should be noted that this is different from the territorial perspective, which addresses the activities taking place within the territory, independent from where the subject resides (Usubiaga & Acosta-Fernández, 2015). The territory principle is used in the Kyoto accounting. In contrast, the residence principle complies with the accounting principles described in the System of National Accounting (SNA) and the System of Environmental Economic Accounting (SEEA) and is common among the input-output practitioners. The main difference between the two principles centres on the treatment of international transport activities carried out by resident and non-resident unit. Usubiaga & Acosta-Fernández (2015) point out that the residence and territorial principles are often used interchangeably. Here we consider production-based emissions which are compiled according to the residence principle. Quantitative differences between consumption, production and territorial accounting perspectives are shown in Appendix Figure 24.

The CBA method considers all environmental impacts resulting from the consumption of goods and services within a country (including private households), irrespective of the geographic location where the production of these goods and services has taken place. Under this principle, all environmental impacts occurring along the chains of production and distribution are allocated to the final consumer. This means that environmental impacts related to the production of imports are taken into account (red box in Figure 2), but those associated with exports are not included.

In this study, we consider both production and consumption perspectives. It should be noted that PBA and CBA accounting principles represent two extremes in the sense that they assume full consumer or full producer responsibility (Lenzen et al., 2007). Several other methods have been proposed in the literature including income-based responsibility; shared-responsibility (Galego and Lenzen, 2005), technology-adjusted responsibility (Kander et al., 2015), value added-based responsibility (Pinero et al., 2018), emission responsibility allotments (Dietzenbacher et al., 2020) and economic benefit shared responsibility (Jakob et al., 2021).

Figure 2 Production- and Consumption-based accounting principles

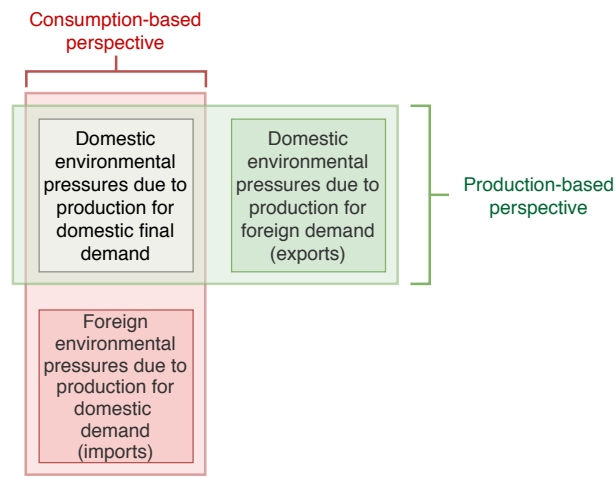

## 2.3. Ethical dimension (normative choices): equity principles and allocation approaches

In the absence of internationally agreed methodology as to how a planetary budget shares should be derived/assessed, scientists and government representatives employed a range of equity principles such as responsibility, capability equality, sovereignty (Höhne et al., 2014) as the basis for effort-sharing approaches and distribution of global Planetary Boundaries. Equity and allocation principles (e.g., cost-effectiveness) commonly discussed in the literature include (Höhne et al., 2014; Lucas et al., 2020; EEA-FOEN 2020):

- *Sovereignty or acquired rights*: all countries have equal right in the use of global collective goods (such as the global atmosphere), and current usage constitutes a 'status quo right'.

- *Equality*: all humans have equal claim to global collective goods (i.e., global commons) such as atmospheric space.

- *Responsibility*: countries that contribute most to the problem should take the largest share of the burden.

- *Capability or capacity*: the greater the capacity to act or pay, the greater the share in global mitigation action. Lucas et al. (2020) notes that the basic need principle (see, e.g., Defila and Di Giuilo 2021) can be considered a special expression of the capability principle, i.e. least capable countries could have a less ambitious reduction effort to secure their basic needs.

- *Cost-effectiveness*: take mitigation action where it is the most cost-effective. Note that this approach is not an equity principle, but is sometimes assumed as an allocation approach. Cost-effectiveness may also be used for *implementing* one of the former principles (e.g., in compensation projects)

Several effort sharing approaches have been derived based on these principles. Typically, these effort sharing approaches include; equal per capita emissions/resource use, contraction and convergence, grandfathering, development rights and ability to pay (van den Berg et al., 2019). A distinction can be made between rights-based and duty-based approaches (den Elzen et al., 2002). Generally (but not always), approaches based on the equity principles *sovereignty* and *equality* establish a right to certain levels of resource use or pollution, while approaches based on *responsibility*, *capability* or *cost-effectiveness* establish a duty to contribute to mitigation.

Lucas and Wilting (2018) and Lucas et al. (2020) examined six distinct allocation approaches. They demonstrate that the share of the assigned planetary boundary to a country level might vary considerably based on the choice of the allocation approach. Indeed, there is no "one size fits all solution" to allocate planetary boundaries. Some approaches might be more/less suitable for specific indicators and different sub-global scales (e.g., countries, sectors). In its simplest form, each approach faces difficulties. Raupach et al. (2014) note that allocating planetary boundaries based on the grandfathering approach would leave many lower-income countries with limited access to energy and development opportunities. On the other hand, the equal per capita share approach would impose high mitigation demands on many high-income countries.

These difficulties have stimulated the development of the "blended" (Raupach et al. 2014), "Triptych" (Phylipsen et al., 1998) and "multi-stage" (den Elzen, 2002) effort-sharing approaches. Despite some differences, these approaches are similar in the sense that they are based on more than one equity principle. For instance, the Triptych approach incorporates most important national circumstances relevant to GHG emission limitation and reduction, such as differences in the standard of living and level of economic development, differences in economic structure and differences in the fuel mix (Phylipsen et al., 1998). This approach makes it possible to calculate in a relatively simple way what individual contribution countries can "reasonably" be expected to make to an overall emission reduction target. The Triptych approach was used as a basis to share the emissions reductions of the first commitment period for the Kyoto Protocol within the EU (Höhne et al. 2014). In general, various blended/mixed approaches have been applied in the carbon (or greenhouse gas) budget literature but not often in the broader planetary boundary framework. According to a recent review by Bjørn (2020), most studies apply equal per capita, grandfathering and economic value-added allocation approaches.

For this study, we have selected four distinct allocation approaches: grandfathering, equal per capita allocation, ability to pay and value-added based approach (for sectoral allocation). Also, we present an example of a blended approach. A summary of the different effort-sharing approaches is shown in Table 1, and each of them is discussed in more detail below. Other options that take into account differences in environmental conditions, e.g., Spain is relatively dry and is likely to have much higher water requirements than less arid regions; or Norway and Switzerland are mountainous, and various infrastructure projects (e.g., building a road/tunnel) are likely to be associated with higher resource use and emissions than similar projects in other countries. Grandfathering approach partially captures these aspects, but other metrics are needed that can better capture these differences. Furthermore, it should be emphasised that in this report, we do not aim to answer the question of which principle or combination of principles is the best founded or most plausible in terms of ethical considerations. Instead, we explore different options and outcomes.

Table 1 Effort sharing approaches and equity principles

| Approach | Equity principle | Parameters |
|---|---|---|
| Equal per capita (EPC) | Equality | Population |
| Grandfathering (GF) | Sovereignty or acquired rights | GHG, $CO_2$, PDFyr, Freshwater use |
| Ability to Pay (AP) | Capability / capacity | Value-added, Employment |
| Value-added (VA) | Utilitarianism | Value-added |
| Blended Approach | Combination | |

### 2.3.1. Equal per capita (EPC) allocation

In the planetary boundary framework studies often rely on the equal per capita allocation approach. The equal per capita share approach is based on the equality principle and reflects the population distribution. Formally this approach can be expressed as:

$$s_i^{epc} = \frac{pop_i}{POP} \tag{1}$$

Where $POP$ is population total for the world; $pop_i$ is population total for country $i$; $s_i^{pop}$ is the share of the allocated quota to country $i$ based on the equal per-capita approach. Besides the equal per capita shares in a given year, the allocation can also be based on the projected future population and/or account for past resource uses (see, e.g. Dao et al. 2018).

### 2.3.2. Grandfathering (GF) allocation

The grandfathering approach (also known as "inertia") reflects the distribution of a country's environmental pressure or impact. Formally this approach can be expressed as:

$$s_i^{gf} = \frac{ep_i}{EP} \tag{2}$$

Where $EP$ is environmental pressure total for the world; $ep_i$ denotes current total environmental pressure for country $i$; $s_i^{gf}$ represents the share of the allocated quota to country $i$ based on the grandfathering approach. The grandfathering approach can be based on current and/or past environmental pressures or impacts. Note that the grandfathering approach that does not account for past resource use and emissions may allocate higher overall budgets to those whose environmental impacts have been overshot in the past. The issue about historical responsibility and how it should be addressed is a matter for debate (Friman and Strandberg, 2014) that involves a range of considerations (e.g., a start year). In this study, we do not consider historical responsibility mainly due to practical issues associated with data availability and associated complexities (e.g., footprints estimates are mostly available since 1995). That said, grandfathering offers several important advantages in terms of simplicity, data availability, accounting for environmental conditions (e.g., water-scarce regions are likely to have higher water demand and it seems reasonable to allocate a higher share of the budget to these regions) and negotiations.

### 2.3.3. Ability to pay (AP) allocation

Ability to pay (AP) is a duty-based approach based on the capability principle. It is closely related to the countries' capacity to contribute to solving the environmental issue. This approach can be expressed in different ways, for instance, some researches (van den Berg et al., 2019; Lucas et al., 2020) have used GDP per capita or the inverse of GDP per capita instead of the aggregate GDP of the country. Here we aim to derive an indicator based on the industry data (such as value-added, employment) which would be inversely related to the ability to pay for environmental reduction. In other words, the higher the indicator value, the lower the share of the assigned budget.

Here we use value-added (VA) per employee (EMP) as an indicator. Formally this approach can be expressed as:

$$s_i^{ap} = \frac{emp_i \left(\frac{va_i}{emp_i}\right)^{-\alpha}}{\sum_i \left[emp_i \left(\frac{va_i}{emp_i}\right)^{-\alpha}\right]} \tag{3}$$

Where $va_i$ and $emp_i$ is value-added and employment for country $i$ with a set to 0.5; $s_i^{ap}$ is the share of the allocated quota to country $i$ based on the ability to pay approach. Often high-income countries and industries such as pharmaceuticals and finance are characterised by relatively high VA/EMP values.

### 2.3.4. Value-added (VA) allocation

Value-added allocation is based on utilitarianism, which aims to maximise total welfare. It can also be viewed as the efficiency principle as it seeks to minimise the costs of reducing environmental impacts (Bretschger, 2013). Formally it can be expressed as

$$s_i^{va} = \frac{va_i}{VA} \tag{4}$$

Where VA is a global value-added and $va_i$ is value-added for country $i$ $s_i^{va}$ is the share of the allocated quota to country $i$ based on the value-added approach. It should be noted that reducing environmental impacts is not exclusively an efficiency but also an equity concern. The efficiency view is concerned with total costs becoming minimal, while the equity view argues that these costs should be distributed fairly (Bretschger, 2013).

### 2.3.5. Blended allocation (BA)

Fairness and efficiency are often considered to be at odds, but an appropriate indicator that considers multiple dimensions can alleviate the conflict. A blended allocation approach is based on a mix of more than one effort sharing approach. This approach is common in the carbon literature, but to best of our knowledge has not been applied in the planetary boundary framework. Raupach et al. (2014), used the blended approach to allocate the global carbon budget among different nations. Bretschger (2013) used a combination of effort sharing approaches to build a "fairness index". A weighting index for fairness, used as the main element to determine fair carbon budgets. This method was applied to build the ETH Climate Calculator (http://www.ccalc.ethz.ch/).

Following Raupach et al. (2014), the share of a particular planetary boundary for country *i* can be obtained as a mixture of EPC and GF allocation:

$$s_i^{ba} = (1-w)\frac{pop_i}{POP} + w\frac{ep_i}{EP} = (1-w)s_i^{epc} + w*s_i^{gf} \qquad (5)$$

Where $w$ is a weighting index (between 0 and 1), and $s_i^{ba}$ is the share of the quota allocated to country $i$ weighted between the equal per capita share and current level of emissions. As shown in Raupach et al. (2014), this concept can be generalised to include additional metrics such as GDP that capture capability to implement required mitigation efforts. Here we extending it by adding the ability to pay ($AP$) parameters as follows:

$$s_i^{ba} = w^{epc}s_i^{epc} + w^{gf}s_i^{gf} + w^{ap}s_i^{ap} \qquad (6)$$

Where $w^{epc}$, $w^{gf}$ and $w^{ap}$ are the weights that determine the importance of each metric, and sum to 1 ($w^{epc} + w^{gf} + w^{ap} = 1$). These weights can be divided evenly or adjusted according to specific criteria. In this study, we use equal weights. One key advantage of this approach is that it avoids complexity and comprises egalitarian, ability, and sovereignty aspects, which often appear necessary for achieving an agreement between multiple stakeholders related to environmental issues.

## 2.4. Combining country-specific shares with Planetary Boundaries

Translating global Planetary Boundaries into sub-global scales (e.g., country, city) requires converting its control variables into a global budget. For instance, for the climate change Planetary Boundary we first define a global level of GHGs emissions for a specific year, consistent with meeting a long-term climate objective (i.e., 1.5°C or 2°C by 2100). After this, we apply one of the effort sharing approaches to determine a country-specific share of a global carbon budget. The same procedure applies to biodiversity and freshwater use boundaries. It should be noted that allocation of the global limits to the national scale can be based on two different types of approaches, yearly budgets or budgets over time to consider the future when required (Dao et al., 2015). For example, $CO_2$ is based on budgets over time (as the impact depends on cumulative emissions), while land-use related biodiversity and water footprint are based on a yearly budget to reflect the nature of the impact.

Country specific planetary boundaries are derived by combining country-specific shares (based on one of the effort sharing approaches presented in section 2.3) and global Planetary Boundary as follows:

$$pb_i^* = s_i^* \times PB \qquad (7)$$

Where $PB$ is a specific Planetary Boundary (e.g., global yearly $CO_2$ budget), $s_i^*$ is the share of the allocated quota to country $i$ based on one of the effort sharing approaches (i.e. * represents:

EPC, GF, AP, BA) and $pb_i^*$ is the quota of that planetary boundary allocated to country *i* based on a specific effort sharing approach for a given year or period. The overall allocation procedure is displayed schematically in Figure 1.

## 2.5. Combining sector and city shares with Planetary Boundaries

In most cases, planetary boundaries can be allocated directly from global to city or industry scales. However, in some cases, this may not be possible. For instance, as noted by Muñoz and Gladek (2017), allocation based on the EPC approach is not possible to implement in the company/industry context. One possibility to incorporate equality principle to the company/industry context is by using a two-stage allocation procedure (see Figure 1). First, the planetary boundaries are allocated to a country level based on the EPC or Blended Approach (e.g., 0.5EPC+0.5GF) approach. In the next step, these country-specific planetary boundaries can be assigned to the industry level based on the chosen approach. Such an approach ensures that a country (and the sectors that comprise it) complies with the globally defined boundaries and also incorporates egalitarian principles.

Recently, Algunaibet et al. (2019) adopted this approach in the design of the USA energy systems. Their approach incorporates planetary boundaries into energy systems models, explicitly linking energy generation with the Earth's ecological limits. The USA power sector share of the safe operating space was quantified from the ratio of the USA population to the global population times the ratio of the US power sector Gross Value Added (GVA) to the GVA of the whole US economy. In other words, the safe operating space was first allocated to a country level based on the EPC approach and then to the industry level based on the VA approach.

Given that the parameters (e.g., grandfathering approach) used to allocate the budget are the same and available at a country and sector level, then there is no difference if the budgets are allocated directly to industries, or first to countries and then to sectors. Taking equation 4 as an example, we can show that the sectoral share of the budget is the same:

$$s_{ij}^{ep} = \frac{ep_{ij}}{ep_i}\frac{ep_i}{EP} = \frac{ep_{ij}}{EP}$$

The total budget for a sector is the same, and all sectors add up to a country total. In the environmental accounting literature, this is known as the additivity property (see, e.g., Rodrigues et al., 2006). It implies that each country's responsibility should be the sum of the responsibility of the sectors (or cities) that compose it. The additivity is also ensured in the two-stage approach.

## 2.6. Modelling impacts using input-output analysis

Input-output analysis is a name given to an analytical framework developed by Leontief (1936), in recognition of which he received a Memorial Nobel Prize in Economic Science in 1973. The basic idea behind the input-output analysis is that a national (or global) economy can be divided into a number of interlinked sectors and whose relationships can be represented in a mathematical matrix.

One of the key strengths of the input-output framework is that it can be augmented with environmental and social satellite accounts, which form an appropriate basis for the analysis of environmental and broader sustainability impacts (Wiedmann et al., 2011). The first environmental extensions for the input-output model were developed in the late 1960s (Miller and Blair, 2009). Leontief (1970) had an interest in ecological economics and proposed one of the key methodological extensions to account for the environmental pollution that has later been applied and developed further by many researchers. In the last two decades, environmentally extended global multi-region input-output (EE-GMRIO) analysis has seen a remarkable increase in methodological progress, quality and quantity of underlying data and policy-relevant applications (Wiedmann et al., 2011). It has been applied to measure and evaluate a range of environmental topics including carbon emissions (e.g., Davis and Caldeira, 2010), GHG emissions, energy use, material use, land use, water consumption (Wood et al., 2018). It is often used by national (e.g., in the UK, Sweden, Switzerland, Germany, the Netherlands) and international organization (e.g., UN SCP-Hotspot Analysis Tool, IRP's Global Resource Outlook, UNEP's Green Economy Progress Measurement Framework) to inform decision making.

The environmentally extended global multi-regional input-output (EE-GMRIO) model can be expressed as:

$$\mathbf{E} = \mathbf{Q}(\mathbf{I} - \mathbf{A})^{-1}\mathbf{Y} = \mathbf{QLY}$$

where $\mathbf{A}$ is the matrix of technical coefficients it is obtained as $\mathbf{A} = \mathbf{Z}\hat{\mathbf{x}}^{-1}$, $\mathbf{x}$ is the vector of output $\mathbf{Z}$ the intermediate demand, $\mathbf{Y}$ represents final demands, $(\mathbf{I} - \mathbf{A})^{-1} = \mathbf{L}$ is the total requirement matrix (often known as the Leontief inverse) representing interdependencies between industries, $\mathbf{I}$ is the identity matrix, and $\mathbf{Q}$ is the direct intensity for a given environmental pressure (e.g., $CO_2$).

Note that $\mathbf{Q}$ is expressed as:

$$\mathbf{Q} = \begin{bmatrix} (\mathbf{q}^1)' & \cdots & 0 & \cdots & 0 \\ \vdots & \ddots & \vdots & \iddots & \vdots \\ 0 & \cdots & (\mathbf{q}^r)' & \cdots & 0 \\ \vdots & \iddots & \vdots & \ddots & \vdots \\ 0 & \cdots & 0 & \cdots & (\mathbf{q}^n)' \end{bmatrix}$$

Given $n$ countries with $m$ industries the $(\mathbf{q}^r)'$ is a $1 \times nm$ row vector of direct intensity for a given environmental pressure in country $r$, it is obtained as $\mathbf{q} = \mathbf{f}\hat{\mathbf{x}}^{-1}$, where $\mathbf{f}$ is the direct environmental pressure occurring in each industry and each country. The $n \times nm$ matrix $\mathbf{Q}$ contains direct environmental intensities for $n$ countries. Note that $\mathbf{Q}$ consists of staked $(\mathbf{q}^r)'$ row vectors, each $(\mathbf{q}^r)'$ contains data only for country $r$ with zeros elsewhere. The $n \times n$ matrix $\mathbf{E}$ gives environmental flows between countries.

Production-based environmental impacts for each of $n$ countries are given by $\mathbf{Ei}_n + \mathbf{f}_{hh}$, consumption-based as $\mathbf{i}_n\mathbf{E} + \mathbf{f}_{hh}$, where $\mathbf{i}_n$ is the $n$-element summation vector consisting of ones, and $\mathbf{f}_{hh}$ is the direct environmental impact from households (e.g., due to heating)

Sectoral production-based emissions are given by $\mathbf{f}$. Sectoral consumption-based emissions that consider impacts in the supply chain of industries are obtained as $\mathbf{qL\widehat{Y}i}_n$. To be consistent with the terminology used for countries, we refer to this as sectoral consumption-based emissions. However, it should be acknowledged that such terminology might not be the most appropriate when used for sectors. The "cradle to shelf" or extended production-based perspective might be used as alternatives.

For the methodology to derive city-specific environmental impacts, the reader is referred to Froemelt et al. (2020).

## 2.7. Selected planetary boundaries and their budgets

### 2.7.1. Climate Change

The climate change PB has been defined in terms of atmospheric concentration of 350ppm $CO_2$ and a maximum increase in global radiative forcing of 1 W/m2. The current $CO_2$ concentration is approximately 410 ppm, transgressing the boundary. In 2015, countries agreed to aim at holding the increase in the global average temperature to well below 2°C above pre-industrial levels and to pursue efforts to limit the temperature increase to 1.5°C above pre-industrial levels by 2100. This boundary also allows capturing other GHG than $CO_2$.

There is an associated carbon budget to keep the temperatures below the predefined levels. However, the question of how much carbon budget left to limit global warming to 1.5°C (or 2.0°C) is bound with uncertainties (Peters, 2018) and depends on a set of important parameters that cover multiple considerations. According to some of the latest estimates from the start of 2020 (until 2100), the remaining carbon budget is 985 $GtCO_2$ to stay below the 2.0°C threshold with a 66% probability (CONSTRAIN, 2019). The remaining carbon budget for staying below the 1.5°C threshold with a 50% (66%) probability is 395 (235) $GtCO_2$. A 2.0°C threshold with a 66% probability as a target translates into a yearly budget of 12.31 $GtCO_2$ until 2100, (assuming a constant share over the period). These estimates are broadly in line with the IPCC (2018) SR1.5 budget of about 420 $GtCO_2$ for a 66% chance of limiting warming to 1.5°C, and about 580 $GtCO_2$ for 50% chance (note that the start year is 2018 for the IPCC and 2020 for the CONSTRAIN budget). In 2019, global emissions were 43 $GtCO_2$ (5.58 $tCO_2$/capita): if the current global level of emissions stay the same global carbon budget would be exhausted within one (for 1.5°C threshold) or two (for 2°C threshold) decades.

### 2.7.2. Biodiversity loss

In contrast to climate change, the impact of land use depends on where it occurs and how it is managed. Therefore, this requires context-specific targets defined at various scales (e.g., regional and global). Furthermore, the allocation of the planetary boundary of biosphere integrity to the sub-global level is not only complex and difficult but also the appropriate measures are a matter for debate in the expert community (Häyhä et al., 2018).

In this study, we apply the potential species loss from land use indicator introduced by Chaudhary et al. (2016), based on Chaudhary et al. (2015) and Verones et al. (2016), to quantify the local (based on ecoregion level) and global damage potential of land use for biodiversity.

We use data on ecoregion protected area and available natural habitat from Dinerstein et al., (2017) to derive local-scale safe operating space. The data is categorised into four groups (and can be viewed online at https://ecoregions2017.appspot.com/): (i) *Half Protected*: more than 50% of the total ecoregion area is protected; (ii) *Nature Could Reach Half*: less than 50% of the total ecoregion area is protected, but the sum of total ecoregion protected and unprotected natural habitat remaining is more than 50%; (iii) *Nature Could Recover*: the total amount of natural habitat remaining and the amount of the total ecoregion that is protected is less than 50% but more than 20%. Ecoregions in this category would require restoration to reach Half Protected because the available habitat outside protected areas plus the existing protected areas is below 50%. (iv) *Nature Imperilled*: the total amount of natural habitat remaining and the total ecoregion protected is less than or equal to 20%. Dinerstein et al., (2017) follows the IUCN protected area definition "*protected area is a clearly defined geographical space, recognized, dedicated and managed, through legal or other effective means, to achieve the long-term conservation of nature with associated ecosystem services and cultural values*" which is broadly compatible with CBD definition of a protected area.

Based on these protection statuses we derive reductions targets for land-use related biodiversity loss: 10% for *Nature Could Reach Half*; 30% for *Nature Could Recover* and 50% for *Nature Imperilled,* no target is set for *Half protected* (it is assumed as "safe"). For instance, if ecoregion is classified as *Nature could recover*, we assume that land-use related biodiversity loss should be 30% lower in this ecoregion.

These targets are distributed across countries based on a specific effort sharing approach (e.g., grandfathering). Effectively, this gives SOS available in each ecoregion for a specific country (in this case, Switzerland). The key idea behind this approach is that ecoregions with a lower share of the protected area and remaining natural habitat are likely to have higher biodiversity loss and thus need more stringent reduction targets.

### 2.7.3. Freshwater use

Steffen et al. (2015) proposed a global freshwater planetary boundary at 4000 km$^3$ of consumptive blue water use. However, as mentioned in section 2.1.2 the discussions are ongoing in the scientific community on how to measure and assess this boundary. Here we evaluate transgression of the freshwater PB against the global limit (i.e., 4000 km$^3$) and also at the watershed level.

Building on the recent work by Bjørn et al. (2020) we define the safe operating space (denoted as LB meaning "local boundary", to be consistent with PB) at the watershed level as:

$$LB_k = MAF_k - (HWC_k + EWR_k + 0.15 * MAF_k)$$

Where $MAF_k$ is the Mean Annual Water Flow (MAF) at watershed $k$; HWC is human water consumption, it represents water withdrawal that does not return into the watershed after use; $EWR_k$ denotes annual Environmental Water Requirements, it can be understood as a fraction of water required to ensure "fair" conditions of aquatic systems with respect to pristine flow (i.e., flow without human intervention). EWR is based on Pastor et al. (2014) and was taken from the AWARE methodology. Overall LB indicator is closely related to the UNEP-SETAC water scarcity indicator AWARE. One key difference is that it incorporates the factor 0.15 * $MAF_k$ to reflect uncertainties in EWR calculations and ensure a precautionary value of $LB_k$ (see SI material in Steffen et al. 2015 for an elaborate discussion of these uncertainties).

We use data from Pfister and Bayer, (2014) and Boulay et al. (2018) to quantify water consumption, HWC, MAF and EWR at the watershed level (see section 4.6.1). For the global boundary, we use blue water consumption data available in EXIOBASE v3.7 satellite extensions.

Additionally, we also assess the water scarcity footprint based on the AWARE method (Boulay et al. 2018), which reports water stress (in $m^3$eq) and can thus be used for assessing supply chain issues. These results are presented as "water stress" in section 4.

**Multiscale Orientation Values for Biodiversity, Climate and Water: A Scientific Input for Science-Based Targets**

Table 2 Global and local-level budgets for three planetary boundaries

| Planetary Boundary | Global Limit(s) | Local Limit(s) | Source |
|---|---|---|---|
| Climate change | Remaining budgets for $CO_2$ from the start of 2020:<br>235 Gt $CO_2$ (1.5°C 66%)<br>395 Gt $CO_2$ (1.5°C 50%)<br>985 Gt $CO_2$ (2.0°C 66%)<br>1315 Gt $CO_2$ (2.0°C 50%)<br><br>*Notes:*<br>The budgets are from the start of 2020, but the analysis is conducted for 2016. Therefore, these budgets are adjusted by adding emissions of the past four years ~168 Gt $CO_2$<br><br>Yearly budgets are obtained by dividing the remaining total budget by 84 (number of years until 2100)<br><br>The budget for GHG emissions is obtained by multiplying the budget for $CO_2$ by 1.25 (This is a rough approximation that has been used in previous studies see, e.g., Willet et al. (2019). | N/A | Source: CONSTRAIN (2019) |
| Biodiversity Loss | n/a | ecoregion PDF * protected area target | *Global Limit:*<br>n/a<br>*Local Limit:*<br>This study |
| Freshwater Use | 4000 km$^3$ | MAF-(HWC+EWR+ 0.15*MAF)<br><br>MAF – mean annual water flow<br>HWC – human water consumption<br>EWR – environmental flow requirements | *Global Limit:*<br>Steffen et al. (2015)<br>*Local Limit:*<br>This study based on Steffen et al. (2015) and Bjørn et al. (2020) |

# 3. Data

## 3.1. Input-Output data and environmental accounts

We use the latest publicly available EXIOBASE database version v3.7 to perform the analysis (Stadler et al., 2019). EXIOBASE v3.7 provides a time series of environmentally extended multi-regional input-output (EE MRIO) tables ranging from 1995 to 2016 for 44 countries (28 EU member plus 16 major economies) and five rest of the world regions (Stadler et al., 2018). EXIOBASE 3 builds upon the previous versions of EXIOBASE by using rectangular supply-use tables (SUT) in a 163 industry by 200 products classification as the main building blocks.

Other key sources used in this study include spatially explicit data from Pfister and Bayer, (2014) and Boulay et al. (2018) on water availability and consumption at the watershed level. Characterization factors for potentially disappearing fraction of species at global and ecoregion level from Chaudhary et al. (2016).

# 4. Results

## 4.1. Global trends and current status

Country specific results for GHG emissions, water stress and freshwater use are presented in Figure 3, Figure 4 and Figure 5, respectively. The results are displayed on a per capita basis for 2016, and the total change from 1995 to 2016 is shown in a lower part of the figure. Note that PBA and CBA averages represent county averages i.e., (CBA per capita in country 1 + CBA per capita in country 2)/2, while global average represent population weighted result i.e., global emissions / global population.

As shown in Figure 3, high-income countries display higher GHG emissions per capita than lower-income countries, regardless of the measure (i.e., PBA or CBA). The growth rates indicate that lower-income countries are catching up. Furthermore, in many high-income countries, GHG emissions have declined over time (negative bars in a lower part of the figure) regardless of the measure. The opposite trend has taken place in lower-income countries. In China, GHG emissions have grown the most (180% for CBA and 170 % for PBA). These trends imply that GHG emissions per capita are converging, i.e., GHG emissions per capita are declining in high-income countries from an initially high level, and increasing in lower-income countries from an initially low level.

Swiss PBA emissions per capita are amongst the lowest of the high-income countries in the sample and close to the global average. In contrast, CBA emissions per capita are about twice the size of PBA emissions and above the global average. Such a significant difference between CBA and PBA is not common. In most other countries, the difference between CBA and PBA is considerably lower. These differences are usually attributed to three factors (see, e.g., Jakob and Marschinski, 2012): differences in the trade balance between countries, specialization (mix of export and import bundles) and different factor intensities (e.g., $CO_2$ per unit of output). Lastly, it should be noted that when expressed on a per capita basis, Swiss PBA emissions have declined by 32% and consumption-based by 17% between 1995 and 2016.

**Multiscale Orientation Values for Biodiversity, Climate and Water: A Scientific Input for Science-Based Targets**

Figure 3 GHG emissions per capita (in 2016) and total change between 1995–2016

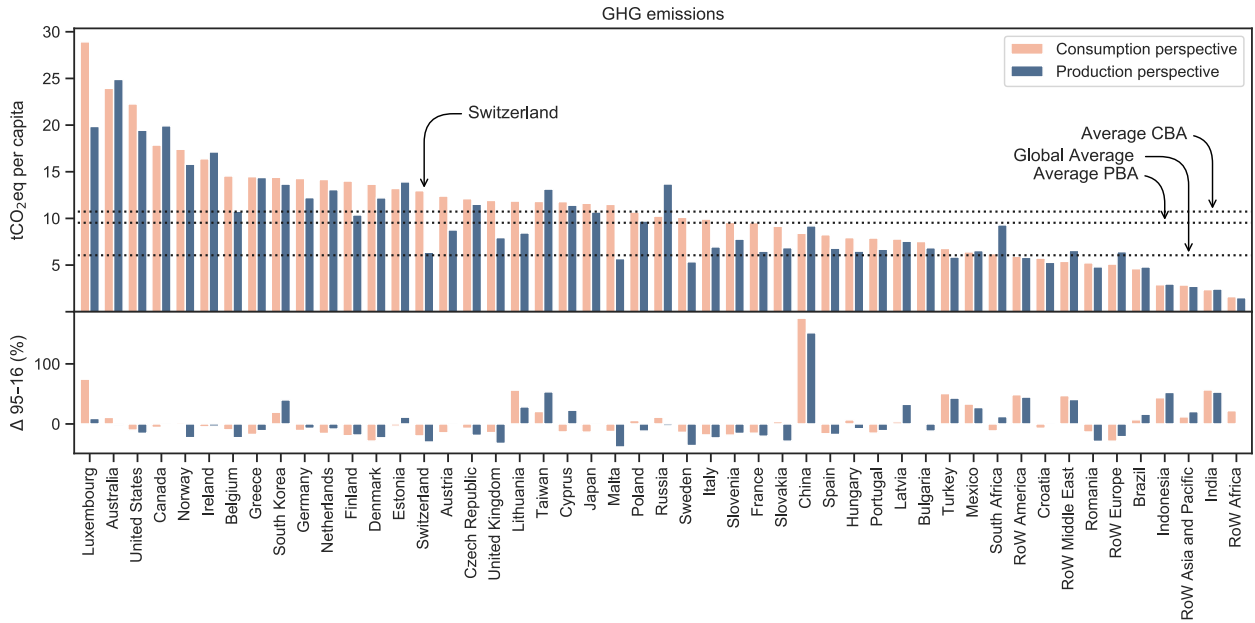

Water stress per capita displayed in Figure 4 show considerable variation across countries and regions. In RoW Middle East, the PBA water stress per capita is considerably above the world average. This is also true, but to a lesser extent, for Bulgaria, and China. For most other countries, the PBA water stress per capita is relatively low. In contrast, the CBA water stress per capita is more evenly distributed across countries and regions.

Figure 4 Water stress per capita (in 2016) and total change between 1995–2016 (AWARE indicator)

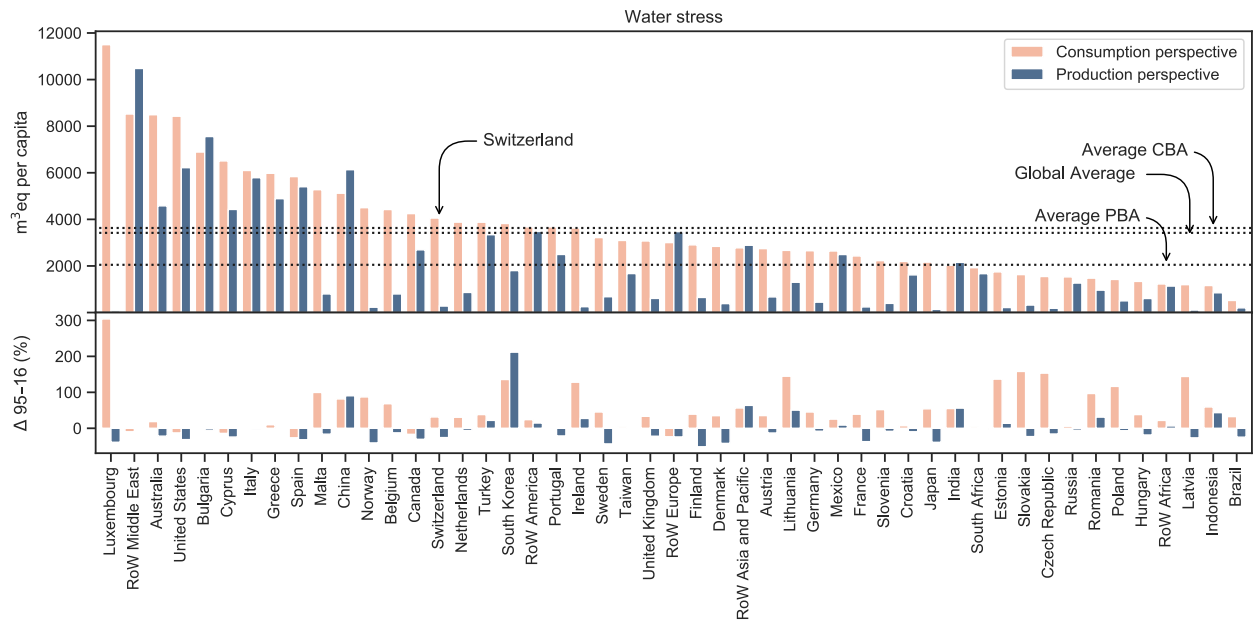

Note that the value obtained for Switzerland (4057 m³eq) is lower than the the 5000 m³eq value reported in Frischknecht et al. (2018).

For Switzerland, water stress from a consumption perspective is similar to the world average, and from a production perspective, the impact is lower than the world average. Water stress per capita had decreased from the production perspective and increased from a consumption perspective during 1995–2016.

Bluewater consumption per capita presented in Figure 5 shows significant variation across countries and regions. Mostly arid and semi-arid countries/regions, such as Spain, Greece and Australia have the highest water consumption per capita from both production and consumption perspectives. Many European countries benefit from economic activities that are less dependent on water consumption resulting in relatively low PBA values for water scarcity. However, often these countries display high CBA water consumption and scarcity per capita, which indicates that imports of agricultural products come from water-scarce regions. Water consumption per capita has decreased in most countries during 1995–2016, mainly due to yield gains for food production.

Swiss PBA freshwater use per capita is very low and substantially below the world average. In contrast, CBA water consumption per capita is considerably higher than the PBA measure and slightly above the world average. Between 1995–2016, CBA and PBA water consumption per capita has declined.

Figure 5 Water consumption per capita (in 2016) and total change between 1995–2016

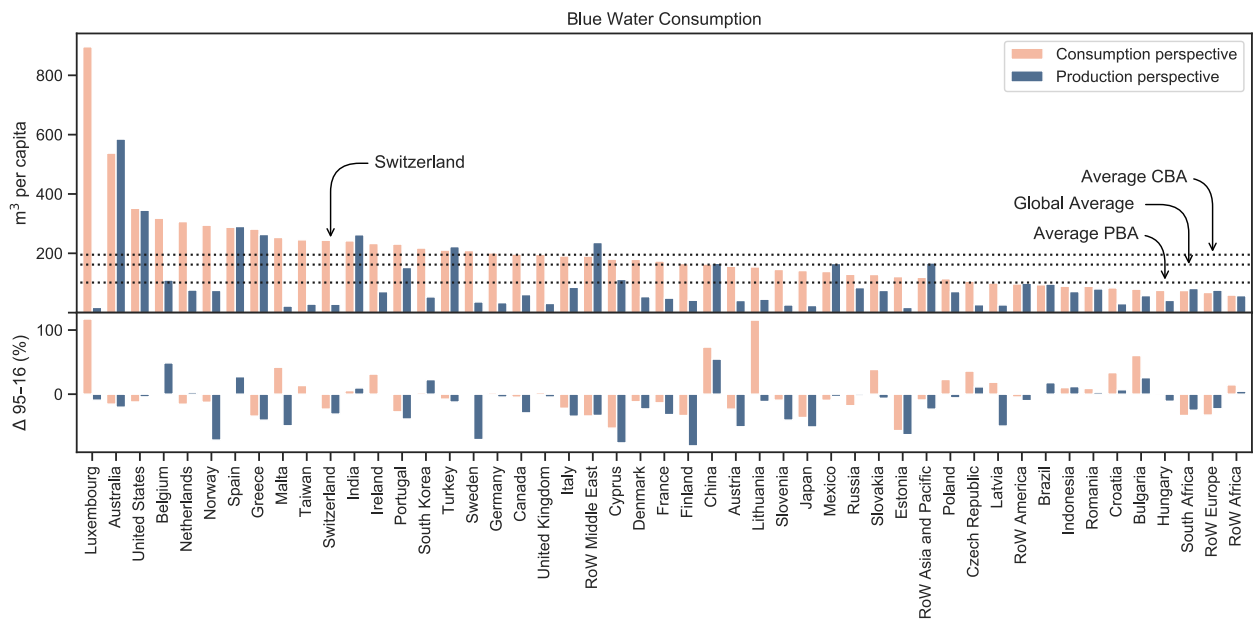

The Water Footprint for Switzerland is 244 m3 per capita. Note that this number cannot be compared with Frischknecht et al. (2018) since the latter (5000 m3-eq) applied the AWARE indicator (accounting for local water scarcities, see above), while the indicator presented here is in m$^3$ without taking into account for watershed-related local scarcities.

In Figure 6 country-specific pressures per capita are normalized by the global average, EXIOBASE sample and the European average. The points below 1 (denoted by the dotted horizontal line) imply that environmental impacts are below global or European average (an example of how to interpret boxplot is shown in Appendix Figure 23). Looking at the Global panel, we find that Swiss impacts are at or below the world average level from the production-based perspective. From a consumption-based standpoint, all impacts are above the global average. Furthermore, from a production-based perspective Swiss GHG, $CO_2$ and water estimates are around the 25$^{th}$ percentile (i.e., Swiss production-based impacts are lower than 75% of the sample). From a consumption perspective, most points fall closer to the 75$^{th}$ percentile.

Comparing Swiss performance within Europe reveals additional insights. First, there is no significant differences in terms of the "Swiss point" (red circle) location in the box plot. Another important observation is that there is room for improvement for the production-based indicators. Most indicators are above the median (50% of European countries) from a consumption perspective.

Figure 6 Switzerland in a global and European perspective, 2016

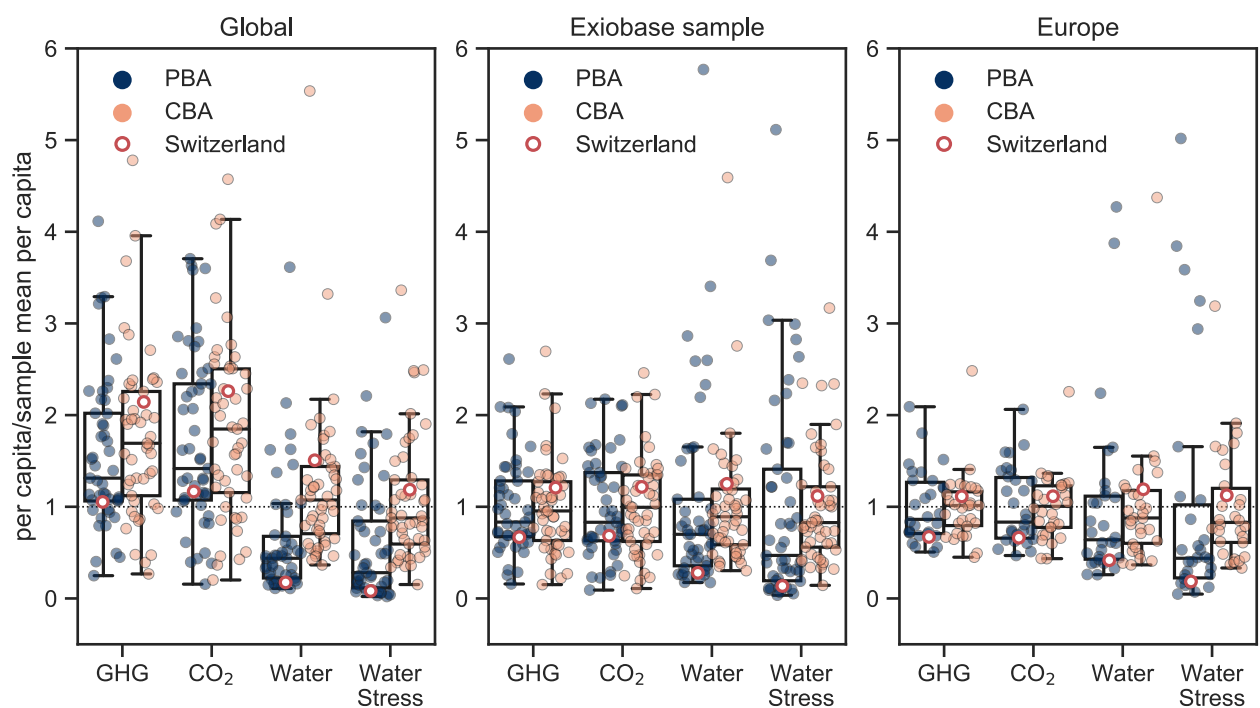

## 4.2. Trends in Switzerland

An overview of trends for the two planetary boundaries (climate and water) for Switzerland are shown in Figure 7. It is evident that for all indicators, the consumption-based (CBA) impacts are higher than those from a production-based (PBA) perspective. Greenhouses gas (GHG) emissions from a production perspective have declined over time, while consumption-based emissions have remained virtually unchanged (note that these are presented in absolute terms). Water stress from a consumption-based perspective has increased over time, while water consumption has declined. Viewed from a production perspective, both water consumption related pressures have remained low.

Figure 7 Trends for two planetary boundaries (climate change and freshwater use) and two indicators for each. Production (PBA) and Consumption (CBA) perspectives, in absolute terms

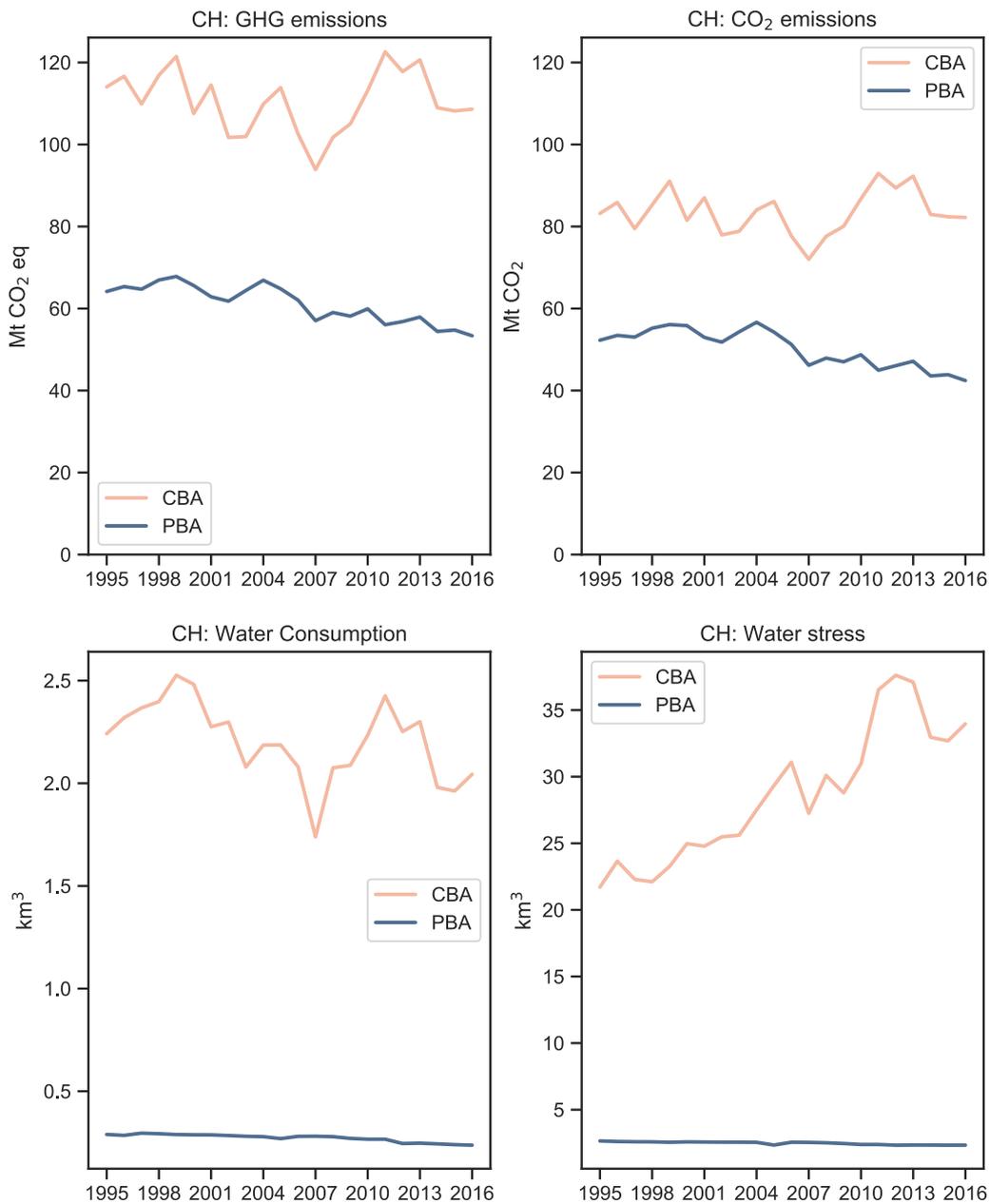

## 4.3. Budget allocation for Switzerland

The shares of the global PBs allocated to Switzerland based on four allocation approaches, and 1.5°C (50%) scenario are shown in Figure 8 for GHG and Figure 9 for $CO_2$ emission (see Figure 25 and Figure 26 in Appendix for different climate targets). Different allocation approaches give very different results. The allocated GHG emissions budget (from consumption perspective) ranges from 0.3 (AP approach) to 2.4 (GF approach) $tCO_2$eq per capita with the average budget of 1.3 $tCO_2$eq per capita (the average is given by BA approach). Note that AP and EPC allocation approaches yield the same budget regardless of the accounting principle (i.e., PBA or CBA). Comparing the allocated budget with GHG emissions in 2016 (denoted by horizontal line), we find that Switzerland appears considerably above the allocated limit for all allocation approaches. It is important to bear in mind that these orientation values apply for 2016. Given global overshot, available budgets for reaching the 1.5°C-target are rapidly declining.

Figure 8 Swiss GHG budget (for 1.5°C 50%) based on five effort sharing approaches, (current level = 2016)

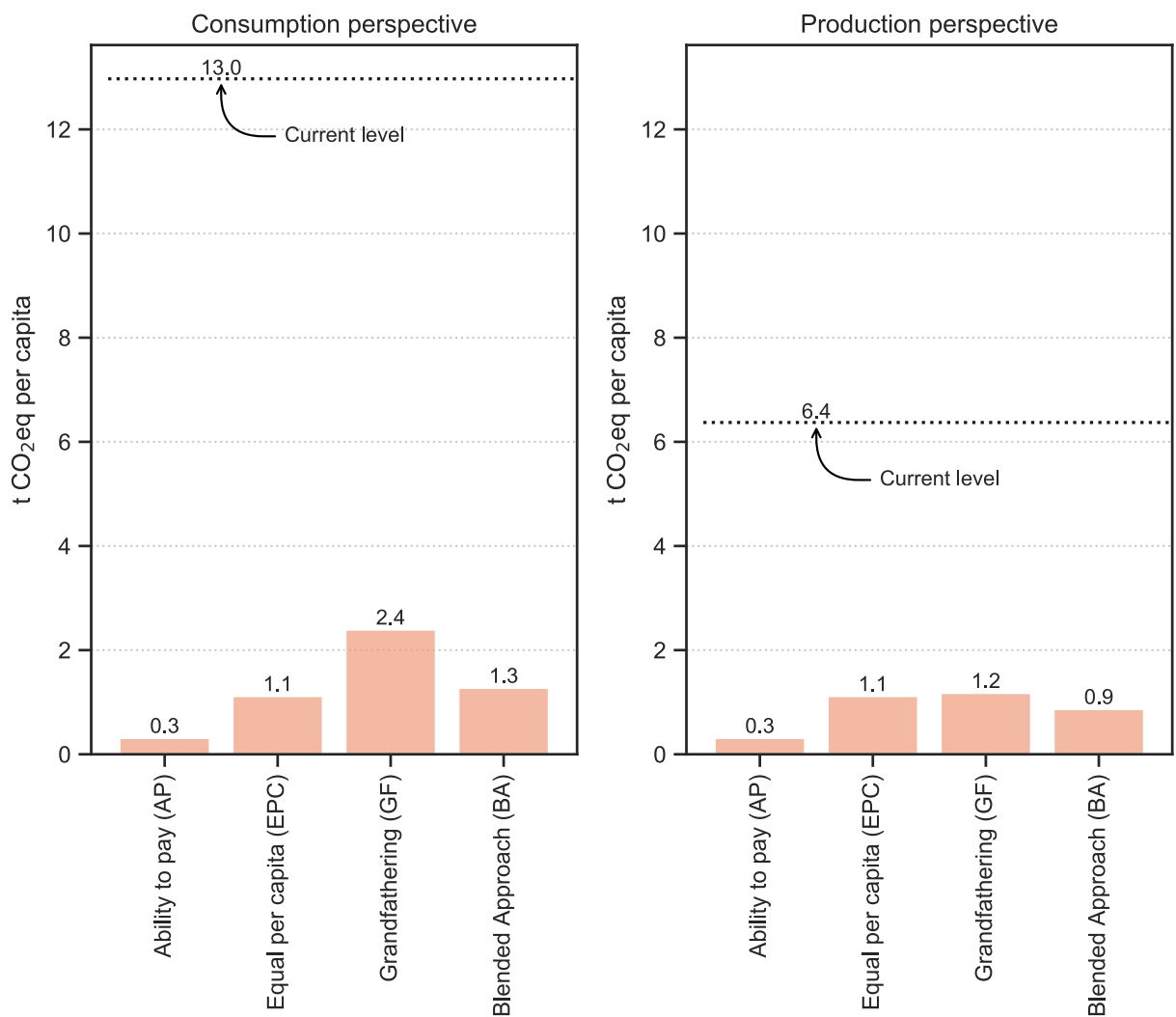

Figure 9 Swiss $CO_2$ budget (for 1.5°C 50%) based on five effort sharing approaches, (current level = 2016)

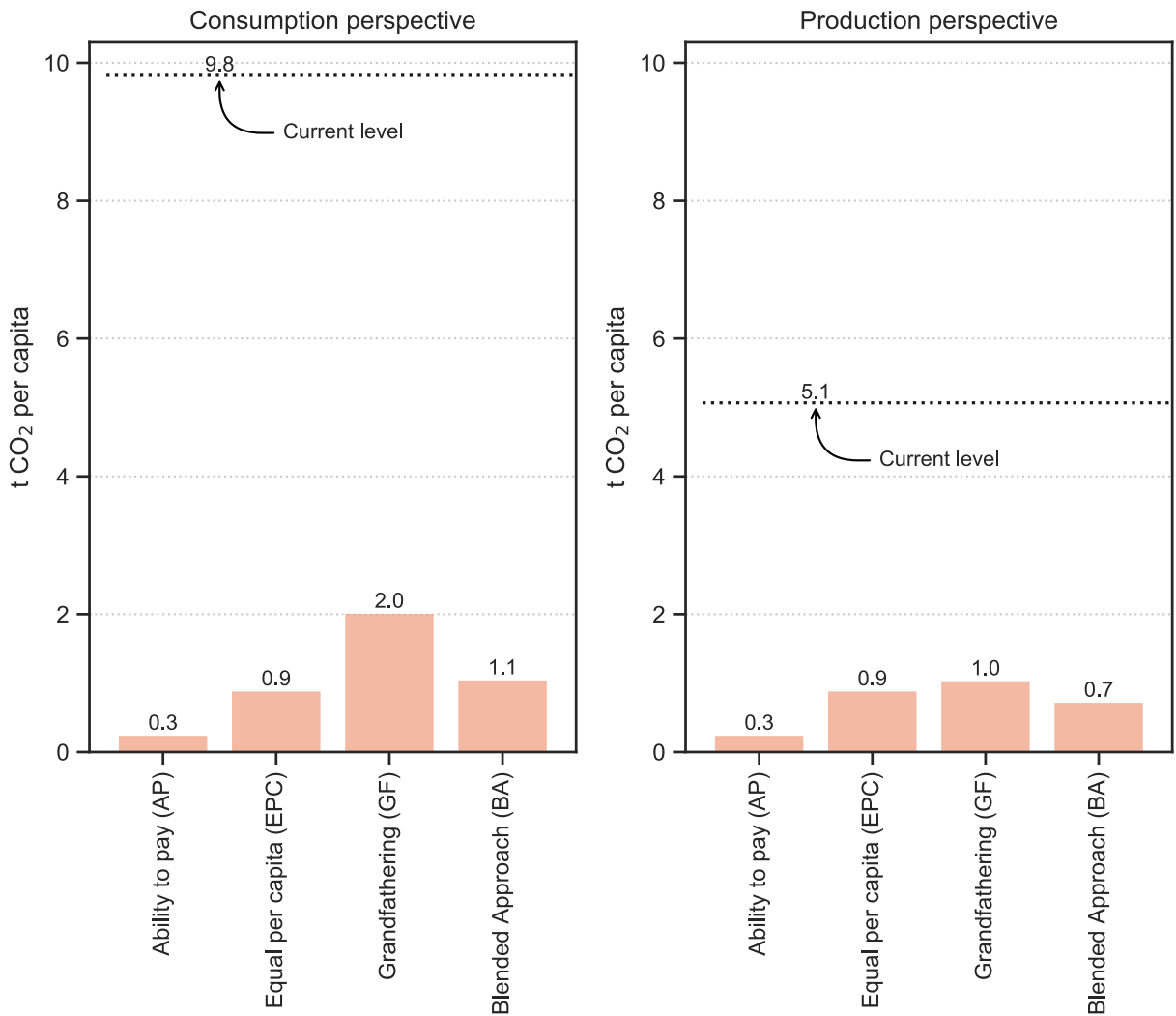

Swiss shares for the freshwater use planetary boundary based on four effort sharing approaches are presented in Figure 10. Interestingly, from a production perspective, the GF approach yields the lowest budget share (93.4 m³ per capita). Overall, the freshwater use planetary is not exceeded by Swiss production or consumption (this is true for all approaches except AP approach from a CBA perspective).

Table 3 and Figure 11 provides a summary of the key results of this chapter. In Figure 11, the transgression of a specific PB is obtained by taking the mean value of the three allocation approaches (EPC, AP and GF). This is equivalent to the Blended Approach. Comparing our results with EEA-FOEN (2020) study for freshwater use (this is the only comparable indicator between the two studies), we find that our results are in close agreement for the min and mean estimates but differ in terms of max values. According to EEA-FOEN (2020) European limit for freshwater use per capita (in m³) ranges between 185 (min), 471 (mean), and 1411(max), respectively our reported values are 150 (min), 496 (mean), and 804 (max).

**Multiscale Orientation Values for Biodiversity, Climate and Water: A Scientific Input for Science-Based Targets**

Figure 10 Swiss Water Consumption budget based on five effort sharing approaches, (current level = 2016)

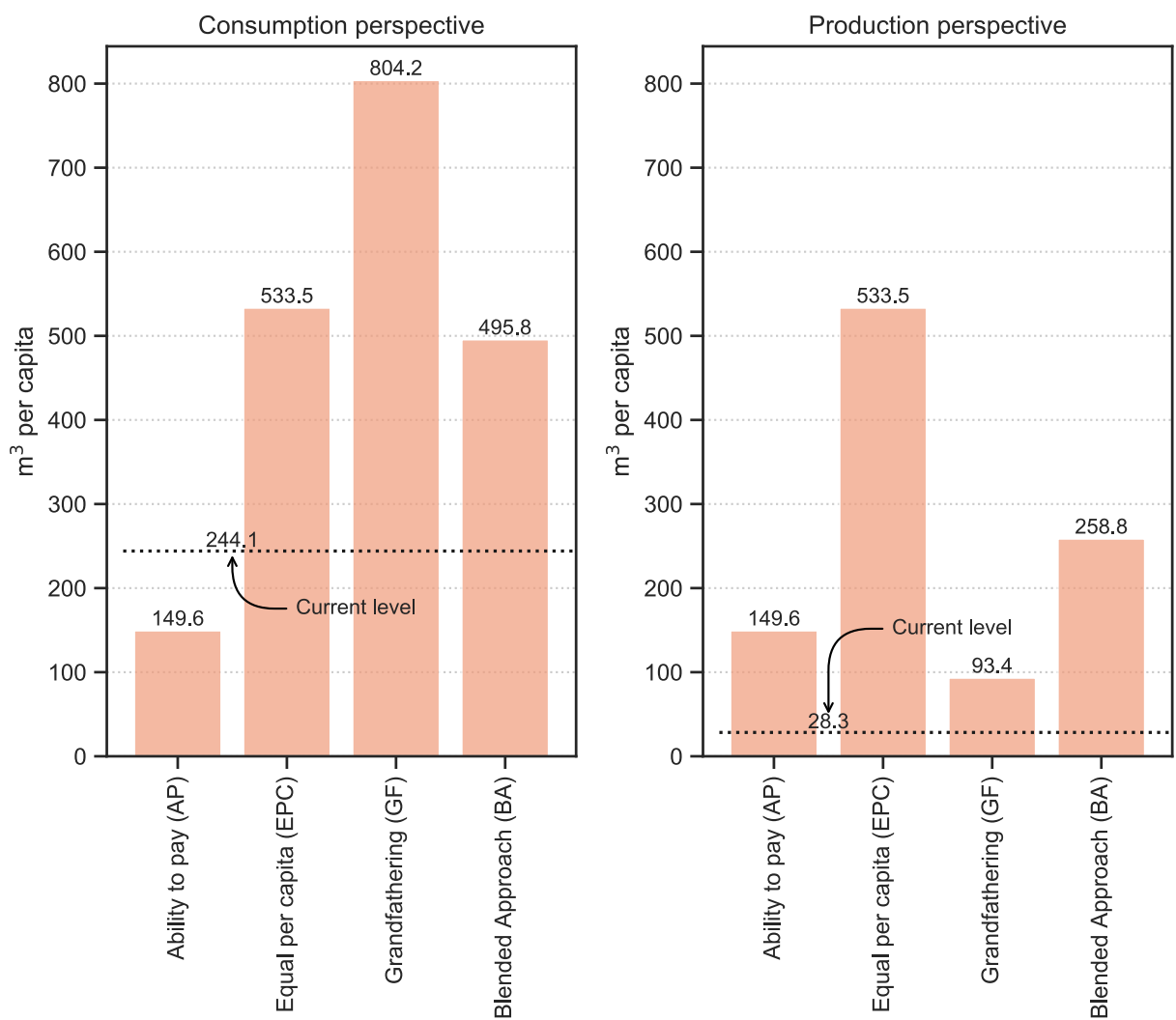

Table 3 Summary of Swiss budgets for three planetary boundaries vs actual impacts, 2016 (per capita)

| | | Consumption perspective | | | | | Production perspective | | | | |
|---|---|---|---|---|---|---|---|---|---|---|---|
| | | Budget | | | | Over/Under shot | Budget | | | | Over/Under shot |
| **Global Boundary** | | Min | Max | Mean | Actual | Actual / Mean | Min | Max | Mean | Actual | Actual / Mean |
| Climate Change (1.5°C 50%) | $tCO_2$ | 0.3 | 2.0 | 1.1 | 9.8 | **10.9** | 0.3 | 1.0 | 0.7 | 5.1 | **7.3** |
| | $tCO_2eq$ | 0.3 | 2.4 | 1.3 | 13.0 | **10.0** | 0.3 | 1.2 | 0.9 | 6.4 | **7.1** |
| Freshwater Use | $m^3$ | 150 | 804 | 496 | 244 | **0.49** | 93 | 534 | 259 | 28 | **0.11** |
| Biodiversity | Global pico PDFyr | 0.6 | 3.0 | 1.9 | n.a. | n.a. | 0.6 | 2.0 | 1.2 | n.a. | n.a. |

**Multiscale Orientation Values for Biodiversity, Climate and Water: A Scientific Input for Science-Based Targets**

Figure 11 Planetary boundaries vs actual performance for Switzerland, 2016

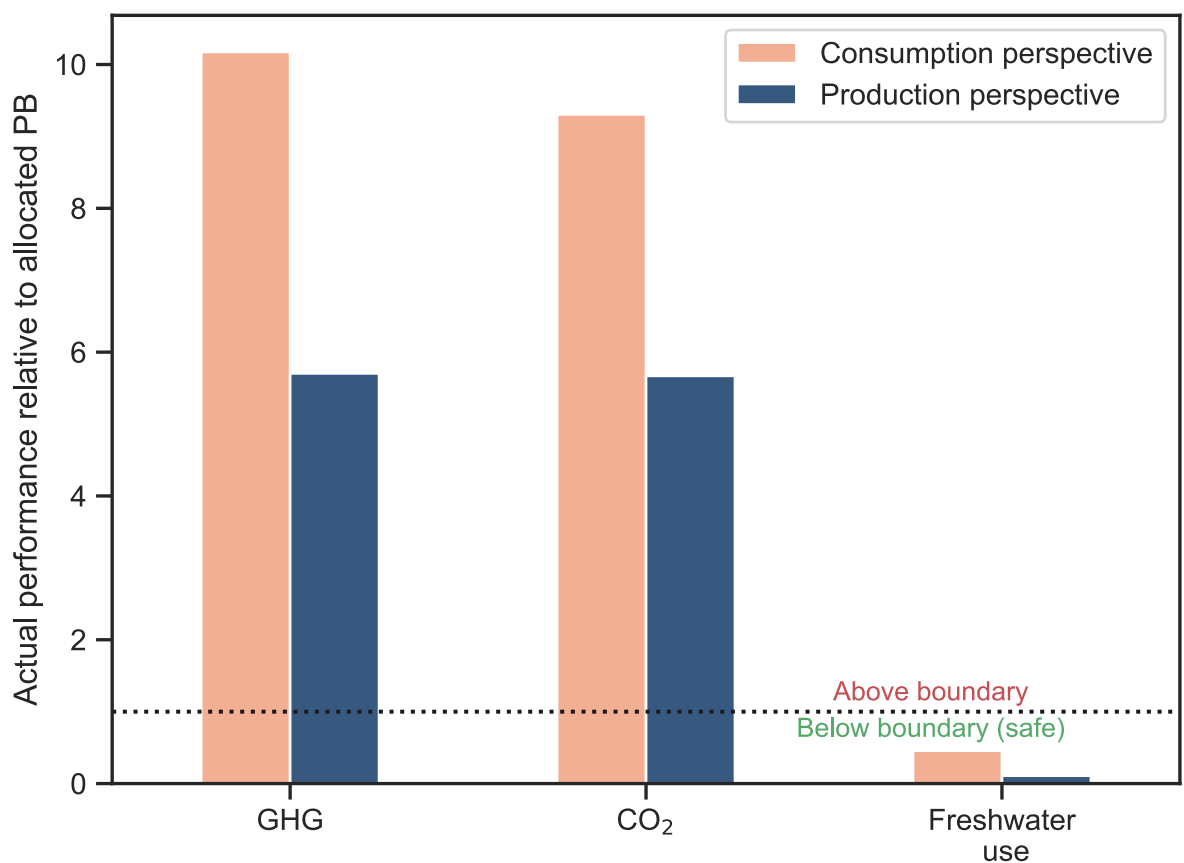

The orientation values for biodiversity loss are shown show in Table 4. Note that we are not presenting actual biodiversity loss results due to possible data uncertainties in the supply chain assessment and observed inconsistencies in the estimates between different studies, which warrant further investigation. The Grandfathering approach (from the consumption perspective) and the Blended Approach, are likely to feature higher uncertainties than the Equal per capita or Availability to pay approach because they are more likely to be influenced by possible uncertainties in trade data. We still present these results for indicating the relevance of allocation selection. The equal per capita value (2 pico PDFyr) is the same as proposed by Frischknecht et al. (2018).

Table 4 Biodiversity loss orientation values for Switzerland (expressed in Global pico PDFyr per capita)

| Allocation approach | Consumption perspective | Production perspective |
|---|---|---|
| Ability to Pay (AP) | 0.6 | 0.6 |
| Equal per capita (EPC) | 2.0 | 2.0 |
| Grandfathering (GF) | 3.0 | 0.9 |
| Blended Approach (BA) | 1.9 | 1.2 |

## 4.4. Sectoral results for Switzerland

In this section, we present results for Swiss industries. The original EXIOBASE v3.7 163 sector classification was aggregated to 33 broad sector categories to allow for more clarity. The GHG and $CO_2$ emissions results are presented in Figure 12 and Figure 13, Freshwater Use in Figure 14. Each figure consists of four panels for each accounting perspective (i.e., PBA and CBA); panel (a) shows the actual environmental impact for respective environmental indicator; panel (b) captures change for a specific indicator between 1995 and 2016; panel (c) compares actual environment pressures against the allocated share of SOS and shows how much it has been over/undershot; panel (d) displays the rate of reduction that is needed to achieve a given target by 2050 or 2100, the reduction rate is based on the blended approach (BA). Note that in panel c some values fall outside the axis limit, this happens because the axis limit is fixed at a specific point (if we don't do this the results get squished), usually this is the case for the ability to pay approach.

Production-based accounting captures direct environmental impacts from owned or controlled sources by a specific sector (e.g., $CO_2$ from driving company vehicles). Consumption-based accounting (we use this term to be consistent with the rest of the document but other names such as "cradle to shelf" or extended-production based approach can be used instead) captures impacts that occur in the entire supply chain of a specific sector. For instance, a car maker in country A buys steel from country B; consumption perspective captures the impact that occurs in country A (car production) and country B (steel production); production perspective captures only what occurs in a car-making sector in country B. Note that the consumption-based approach does not capture impacts associated with the production of goods for the intermediate use (i.e., input into the production of another product). This is done to avoid double counting. Suppose goods are destinated for the intermediate use. In that case, they will be captured by the industry that sells a specific product for final consumption (i.e., a car manufacturer who sells a car for the final consumer will be attributed with the impacts that occurred in the steelmaking industry).

The results for GHG and $CO_2$ emissions presented in Figure 12 and Figure 13 reveal several important insights. For most sectors, consumption-based emissions are above the allocated (most points in panel c fall in red area) budget regardless of the allocation approach. On the other hand, from a production-perspective several sectors in Switzerland are in line with the allocated budget, but this depends largely on the allocation approach. Furthermore, most important sectors (in terms of impact) exceed the budget regardless of the allocation approach. Panel b shows that in many sectors emissions have declined between 1995 and 2016, but in several sectors (especially those where emissions are already high) emissions have increased considerably.

The results also show that the range of the required reduction rates varies considerably for many industries. The reason for this is that different allocation methods favour a different type of industries. The GF allocation approach will favour sectors associated with higher emissions (either directly or indirectly) and allocate the same reduction target to all sectors. In contrast, the VA approach will favour industries that generate more value-added such as finance and pharmaceuticals. This implies that for some industries, the allocation method will play a crucial role in determining whether the industry is within the safe operating space, while for others, it will be is less important. The wide range of allocated budget occurs because Switzerland is an outlier in terms of value-added. Sectors characterized by high value-added will get high budget following VA allocation approach, but very little budget following the Ability to Pay (AP) allocation approach. Other countries/sectors are outliers in other dimensions. For instance, in EEA-FOEN (2020) study, one allocation approach was based on land area. Such an approach would favour large countries with low population density such as Sweden and would be unfavourable for smaller densely populated countries such as the Netherlands. Different indicators will favour different countries/sectors, using multiple indicators and taking their median or mean value can help alleviate these issues.

The freshwater use planetary boundary has not been crossed on the aggregate country level, and as shown in Figure 14, this is also the case at the disaggregated sectoral level. Only in a few instances, sectoral budgets have been exceeded. This happens when the ability to pay approach is used to allocate safe operating space (red dots represent these sectors in the red shaded area of panel c). Panel b shows that from a consumption perspective freshwater use has increased considerably in Food and Beverages and Tobacco as well as Chemicals and Chemical product industries. From a production perspective, the biggest increases are seen in industries that account for a small share of total freshwater consumption (e.g., Public administration and defence). Industries that account for a large share of total freshwater use, (e.g., Electricity, Gas and Water Supply; and Agriculture, Hunting and Forestry), have seen little change in their freshwater use between 1995 and 2016.

**Multiscale Orientation Values for Biodiversity, Climate and Water: A Scientific Input for Science-Based Targets**

Figure 12 Swiss sectoral GHG emissions and allocated budgets, 2016

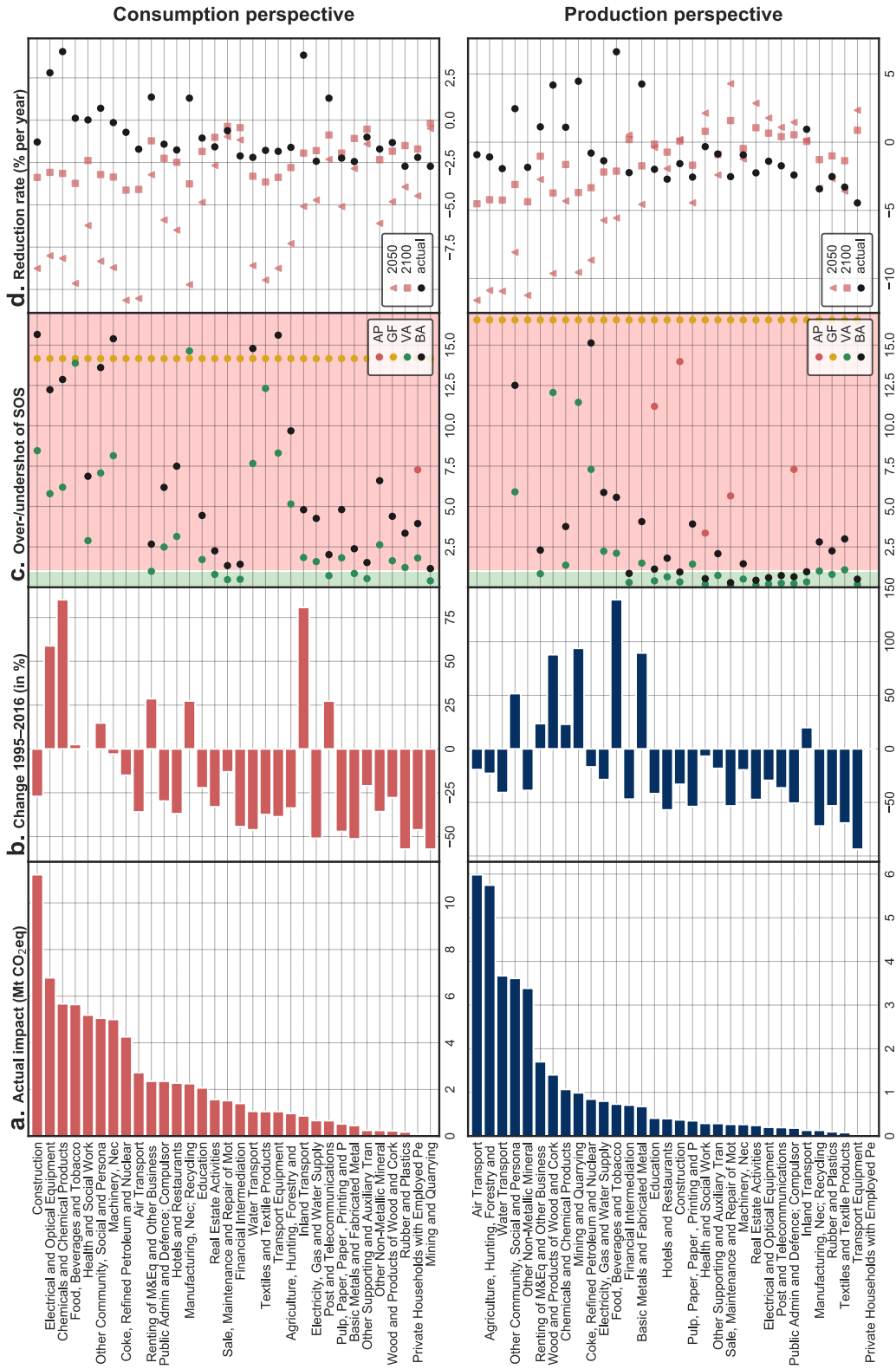

Panel a: Actual impact by a corresponding sector in 2016. Panel b: Change between 1995 and 2016 in %, obtained as (end_year-start_year)/start_year * 100. Panel c: Over/Undershoot of allocated SOS in 2016, obtained as allocated_SOS_x/actual_impact, for allocated_SOS_x, x is replaced by Ability to pay (AP), Grandfathering (GF), Value Added (VA) or Blended Allocation (BA). Area shaded in red denotes overshoot area and green area represents SOS, for clarity panel c only displays values between 0 and 17, global limit is set 6,7 Gt CO₂eq per year (consistent with 1,5°C 50% target). Panel d: Yearly reduction rate to achieve a target by 2050, 2100, the reduction rate is based on the blended approach (BA), actual denotes yearly growth rate between 1995 and 2016.

**Multiscale Orientation Values for Biodiversity, Climate and Water: A Scientific Input for Science-Based Targets**

Figure 13 Swiss sectoral CO$_2$ emissions and allocated budgets, 2016

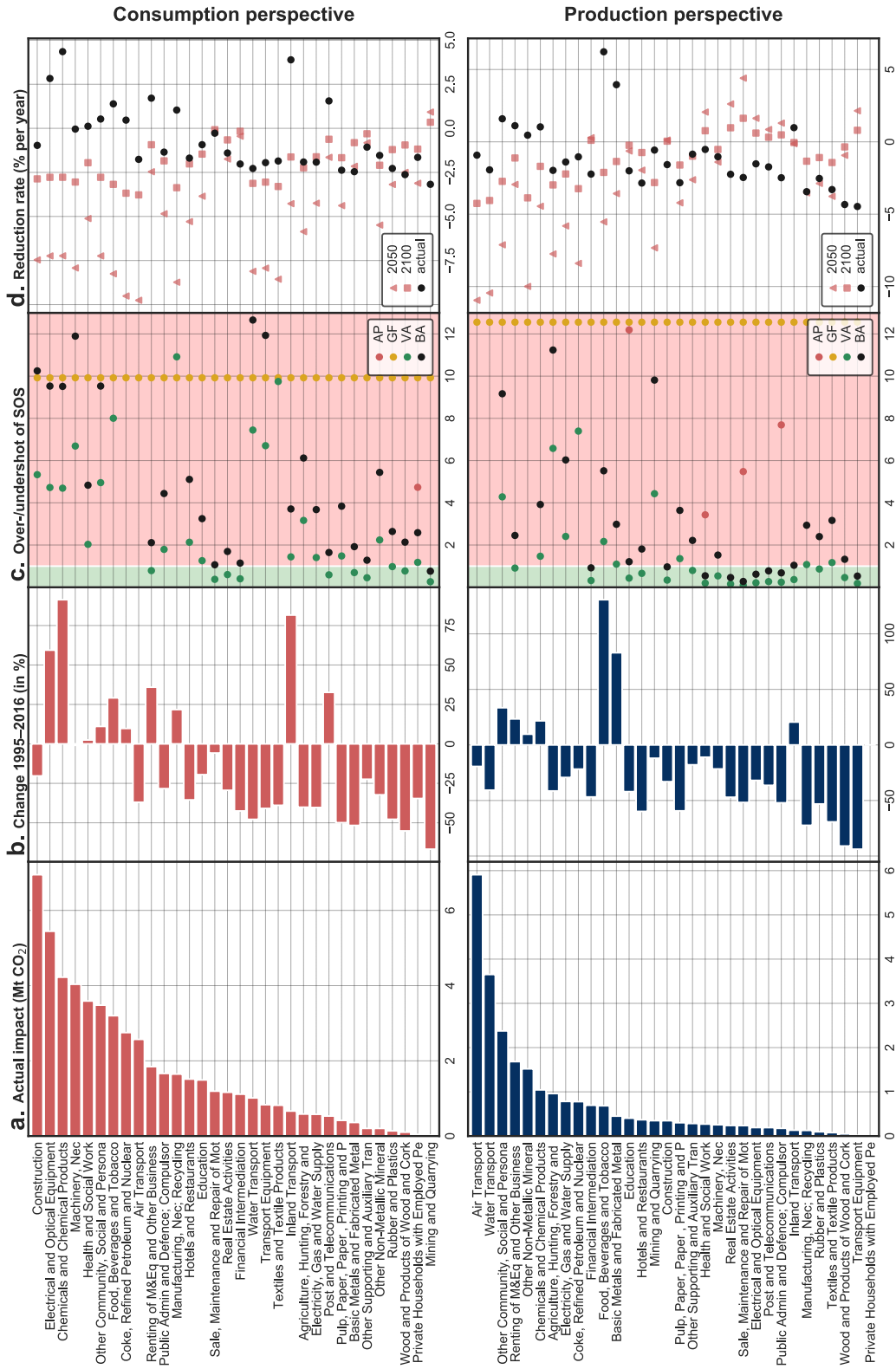

Panel a: Actual impact by a corresponding sector in 2016, Panel b: Change between 1995 and 2016 in %, Panel c: Over/Undershoot of allocated SOS in 2016, obtained as (end_year-start_year)/start_year * 100, Panel c: Over/Undershoot of allocated SOS in 2016, obtained as allocated_SOS_x/actual_impact_for allocated_SOS_x, x is replaced by Ability to pay (AP), Grandfathering (GF), Value Added (VA) or Blended Allocation (BA). Area shaded in red denotes overshoot area and green area represents SOS, for clarity panel c only displays values between 0 and 13, global limit is set at 6.7 Gt CO$_2$ per year (consistent with 1.5°C 50% target), Panel d: Yearly reduction rate to achieve a target by 2050, 2100, the reduction rate is based on the blended approach (BA), actual denotes yearly growth rate between 1995 and 2016.

**Multiscale Orientation Values for Biodiversity, Climate and Water: A Scientific Input for Science-Based Targets**

Figure 14 Swiss sectoral freshwater use and allocated budgets, 2016

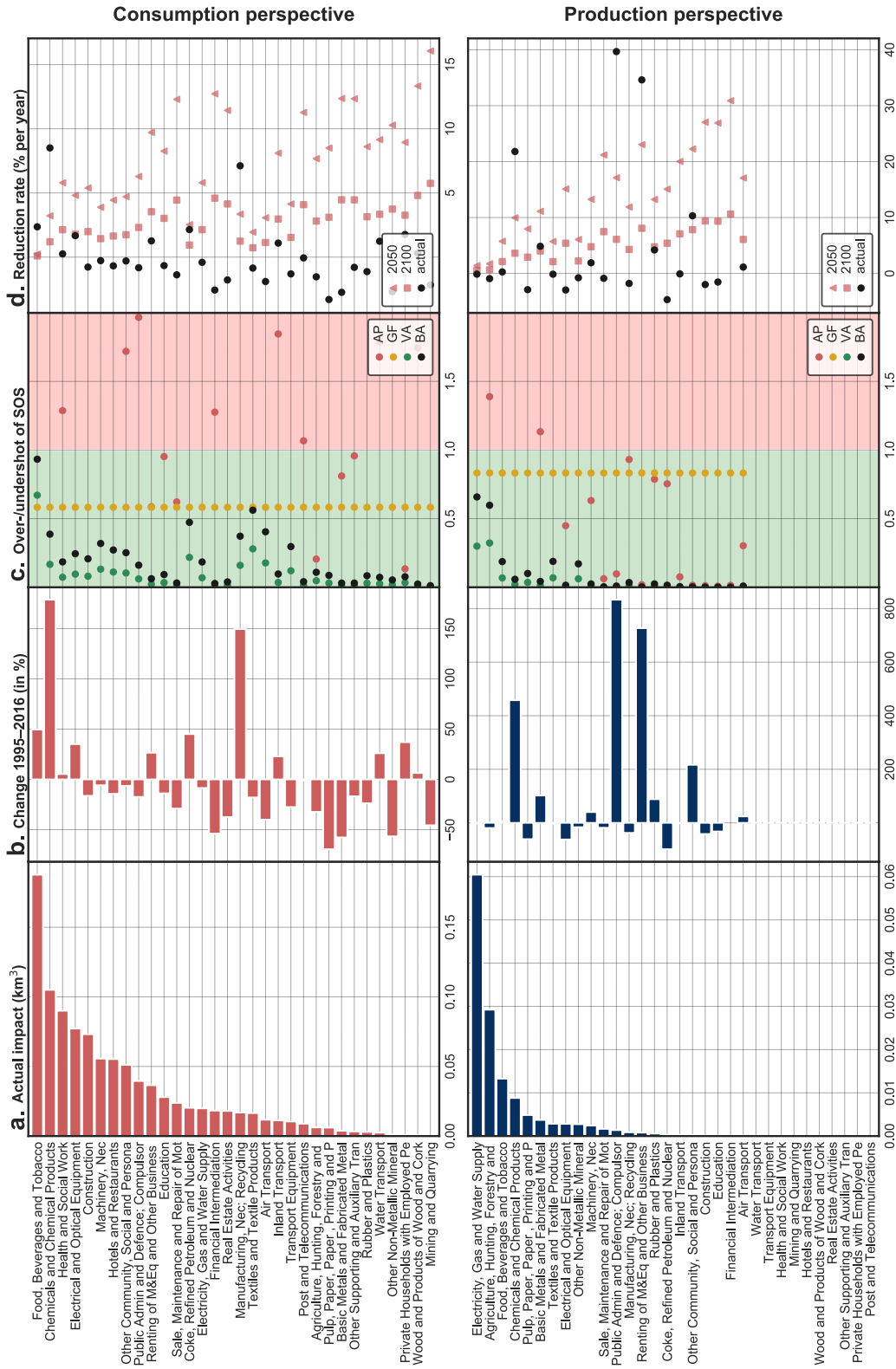

Panel a: Actual impact by a corresponding sector in 2016. Panel b: Change between 1995 and 2016 in %, obtained as (end_year-start_year)/start_year * 100. Panel c: Over/Undershoot of allocated SOS in 2016, obtained as allocated_SOS_x, x is replaced by Ability to pay (AP), Grandfathering (GF), Value Added (VA) or Blended Allocation (BA). Area shaded in red denotes overshoot area and green area represents SOS, for clarity panel c only displays values between 0 and 2, global limit is set at 4000 km³. Panel d: Yearly reduction rate to achieve a target by 2050, 2100, the reduction rate is based on the blended approach (BA), actual denotes yearly growth rate between 1995 and 2016.

## 4.5. City-scale results for Switzerland

Local authorities are increasingly acknowledged to play an important role to curb adverse environmental impacts (see, e.g., Dawkins et al., 2019; Harris et al., 2020). They are close to their inhabitants and often directly provide essential infrastructure, such as energy supply systems, roads or waste disposal schemes. This renders them particularly well-positioned to incentivize green production systems or guide households towards more sustainable consumption behaviours within their jurisdiction (Adua & Clark, 2019; O'Rourke & Lollo, 2015; Tukker et al., 2010). Therefore, operationalising planetary boundaries to a municipal scale is of high importance. A recent example is the city of Zurich. In its environmental report, it has assessed its performance against planetary boundaries (Stadt Zürich, 2020). Regionalized data is an important prerequisite for deriving effective policies targeting the specific situation of a municipality (see e.g., Fry et al., 2018; Horta & Keirstead, 2017; Mirabella et al., 2019; Moran et al., 2018).

To support municipalities in this regard and avoid the cumbersome collection of local data, Froemelt and colleagues (2020) developed a spatially resolved bottom-up model that can combine various large national datasets and tailor this data to the circumstances of a certain municipality. By using machine-learning techniques, the model interlinks three large-scale sub-models a physically-based building energy model (Buffat et al., 2017), a data-driven household consumption model (Froemelt et al., 2018) and an agent-based transport simulation (Hirschberg et al., 2016) and is then able to predict a detailed consumption profile for each actual household in a certain region. In view of the environmental importance of household consumption as a main driver of the economy, a focus on households is legitimate. For instance, Ivanova and colleagues (2016) estimate the households' shares to be 65% of global greenhouse gas emissions and 50% to 80% of total land, material, and water use. However, the model results from (Froemelt et al., 2020) presented in the current section are not fully comparable with other sections of the report. Not only will governmental consumption and other final demand categories are excluded, the model of Froemelt et al. (2020) also comes with an own comprehensive hybrid life cycle assessment framework that combines EXIOBASE v2.2 (Tukker et al., 2013; Wood et al., 2015) and ecoinvent 3.3 (Wernet et al., 2016). Further note that the current section focuses on greenhouse gases only.

Figure 15 presents the absolute as well as the per-capita life cycle greenhouse gas emissions caused by household consumption for all municipalities in Switzerland. While large differences can be observed for the absolute values, the per-capita emissions are more balanced. As can be deduced from Figure 16, no municipality shows household-induced GHG footprints that are below the planetary boundaries and all of them are outside the safe operating space.

Figure 15 Total household-induced GHG footprint for all municipalities of Switzerland

Note: Figures are based on the model by Froemelt et al. (2020). (a) Absolute values of life cycle GHG; (b) Per-capita GHG footprints. The city of St. Gallen is highlighted due to its use as a case study in the report.

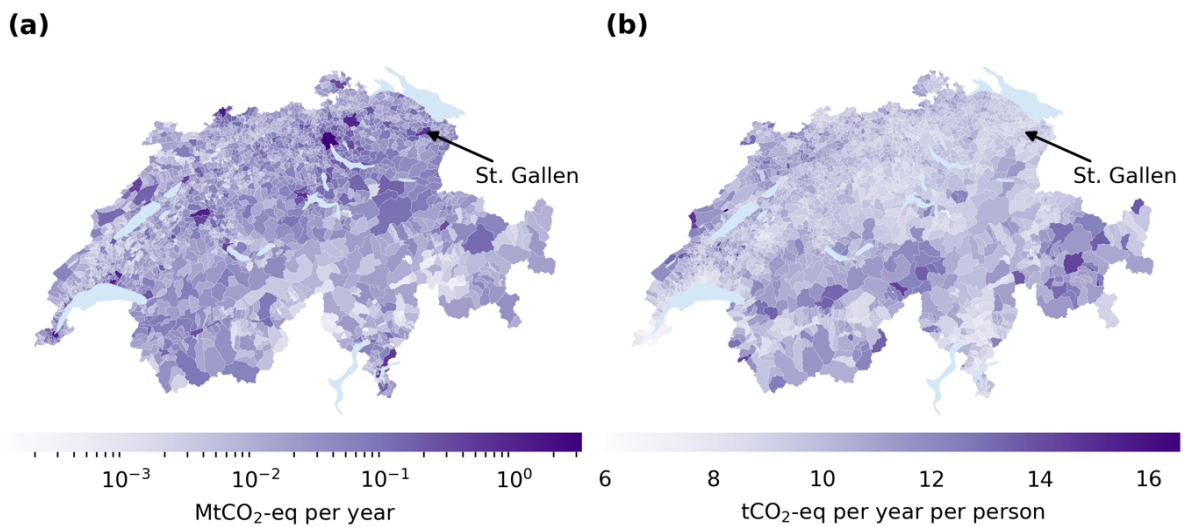

Figure 16 Consumption-based GHG emissions relative to allocated PBs for all municipalities of Switzerland

Note: The budget chosen for this figure is for the 1.5°C (50%) target and the allocation principle corresponds to the blended approach. The city of St. Gallen is highlighted due to its use as a case study in the report. The values in the figure are sorted from the lowest to the highest. [PB=planetary boundaries]

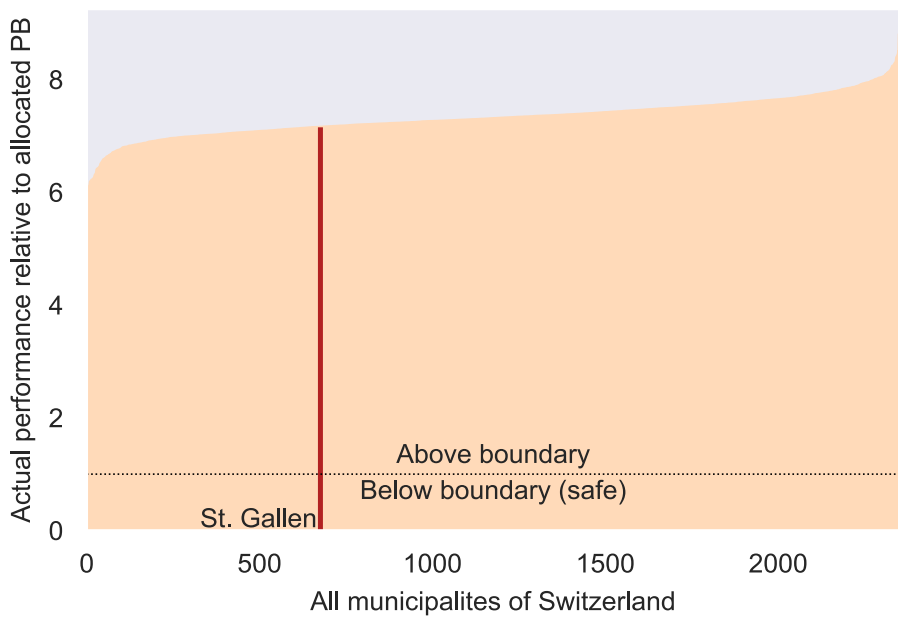

To provide more insights into the GHG footprint of a specific municipality, Figure 17 presents the results of Froemelt et al. (2020)'s model for the city of St. Gallen as a case study. The figure is meant as a hotspot-screening analysis to quickly grasp the situation in the municipality. For this purpose, it aggregates the detailed LCA-results of about 200 different consumption categories for each household in St. Gallen to main consumption domains and at municipal level. However, to preserve some information about household variability, we inserted the interquartile range of per-capita footprints of households as well as Lorenz-curves and Gini-coefficients. Lorenz curves indicate inequality in GHG footprints. The further the Lorenz curve is from the line of equality (diagonal line), the greater the GHG disparity. Gini coefficient measures these inequalities, 0 implies perfect equality and 1 perfect inequality, i.e., the higher the Gini index the further away is the Lorenz curve from the 45-degree line. Figure 17a shows similar per-capita GHG footprints for both Switzerland and St. Gallen. Compared to the Swiss average footprint, St. Gallen features a higher contribution of housing-induced emissions, but a smaller share of transport-caused impacts (see Figure 17a). Housing and transport are also the two most important consumption domains in absolute terms. This becomes further obvious in Figure 17b. Here, the interquartile ranges also show the large variability among households in these two consumption areas. With a Gini-coefficient of 0.53, the emissions induced by transport demand are highly unequally distributed among the households in St. Gallen. Thereby, 17% of the households in St. Gallen are responsible for 50% of the life cycle greenhouse gas emissions caused by mobility needs.

Froemelt et al. (2020) found average per-capita income, population density, the age of the building stock, composition of household types and the access to public transport services as possible drivers for municipal carbon footprints. Figure 18 puts St. Gallen in comparison with all other Swiss municipalities regarding some of the most important potential carbon footprint drivers.

Being classified as "densely populated" by DEGURBA (Eurostat, 2019) should generally decrease the per-capita footprint of St. Gallen. However, the high modeled average income counteracts this tendency and tends to increase municipal carbon footprints in various consumption categories. The large share of buildings built before 1919 in the building stock might be the reason for the higher housing-induced emissions compared to the Swiss average, while the low share of persons living in an area with poor access to public transport is likely to cause lower mobility emissions than in other municipalities. These opposing effects of different variables were already found in Froemelt et al., (2020) on a nationwide level and emphasize the importance of a database that is able to capture regional distinctions when it comes to developing effective policies.

Finally, Figure 17c confronts the total emissions of St. Gallen with three planetary boundary downscaling approaches (equal per capita, grandfathering and blended approach) for the global budget of the 1.5°C (50%) scenario. As could be expected from Figure 16, all three limits are highly exceeded.

**Multiscale Orientation Values for Biodiversity, Climate and Water: A Scientific Input for Science-Based Targets**

Figure 17 Hotspot-screening analysis for the city of St. Gallen

Notes: (a) Comparison of household consumption-induced carbon footprint compositions of St. Gallen and Switzerland. (b) Comparison of per-capita carbon footprints of St. Gallen and Switzerland subdivided into main consumption areas. The error bars show the interquartile range of households (left end: 25%-percentile household; right end: 75%-percentile household). Lorenz-curves and the Gini-coefficient add information about the variability and inequality of life cycle emissions among the households in St. Gallen (Gini scales from 0=perfect equality to 1=highest possible inequality). (c) Juxtaposition of total, absolute life cycle emissions of St. Gallen with allocated planetary boundaries for the 1.5°C (50%) scenario. [GHG=greenhouse gas emissions; HH=households]

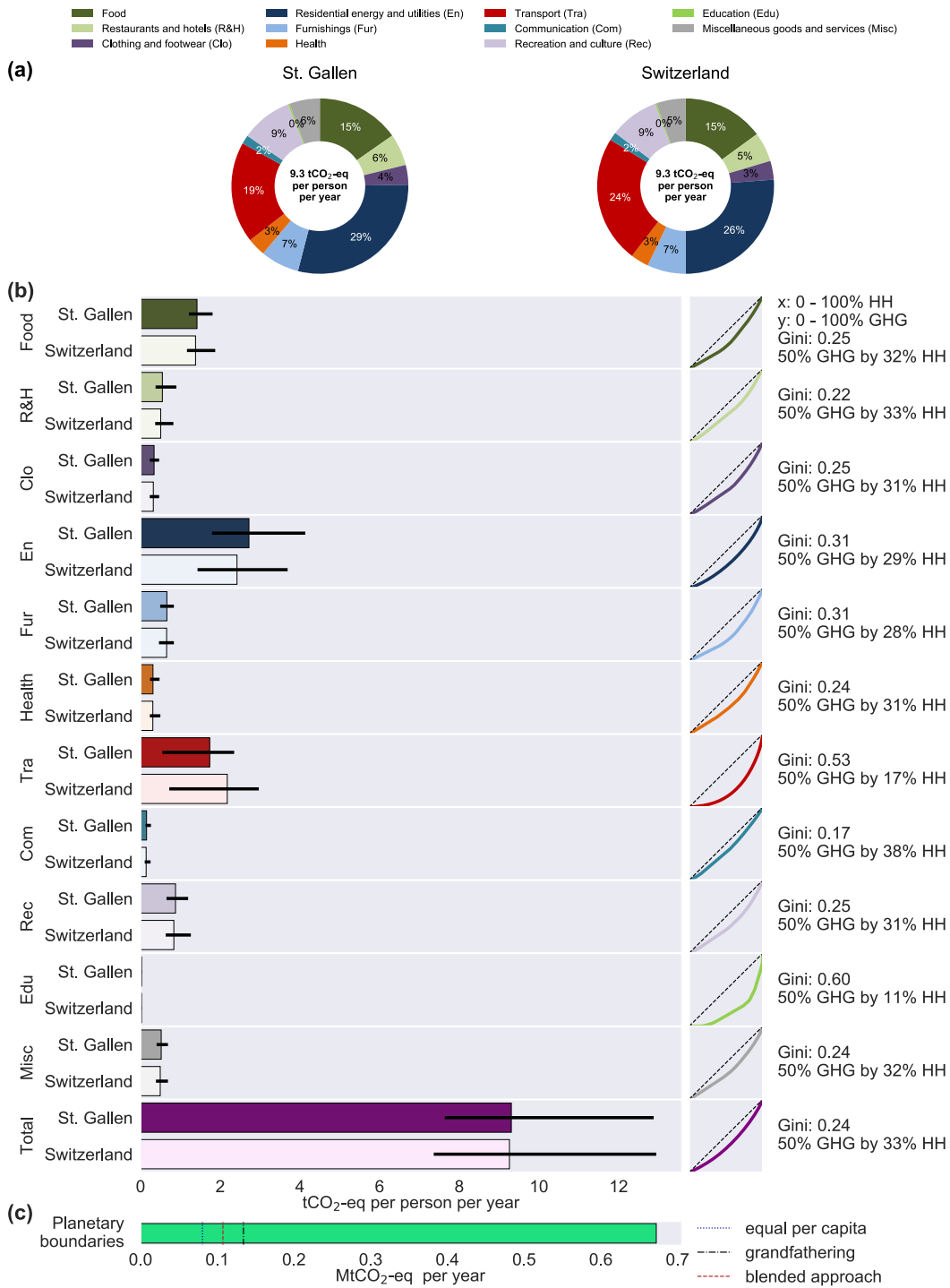

Figure 18 Drivers of municipal GHG footprints for Switzerland

Notes: Violinplots for Switzerland-wide municipal statistics that were identified to be potential drivers of municipal GHG footprints in (Froemelt et al., 2020). The "fattiness" of the graphic indicates frequency (share of municipalities) In the middle of each violinplot, a miniature boxplot is depicted. The dashed red line indicates the position of St. Gallen. The three characteristics affect municipal GHG footprints as follows: the higher the modelled income, the higher the carbon footprint; the higher the share of buildings built before 1919, the higher the carbon footprint; the higher the share of persons living in areas with poor access to public transportation (PT), the higher the carbon footprint. "Poor access to PT" is defined as an area with poorest public transport services (category "unclassified") according to the classification of the Federal Office for Spatial Development (ARE (Bundesamt für Raumentwicklung), 2017).

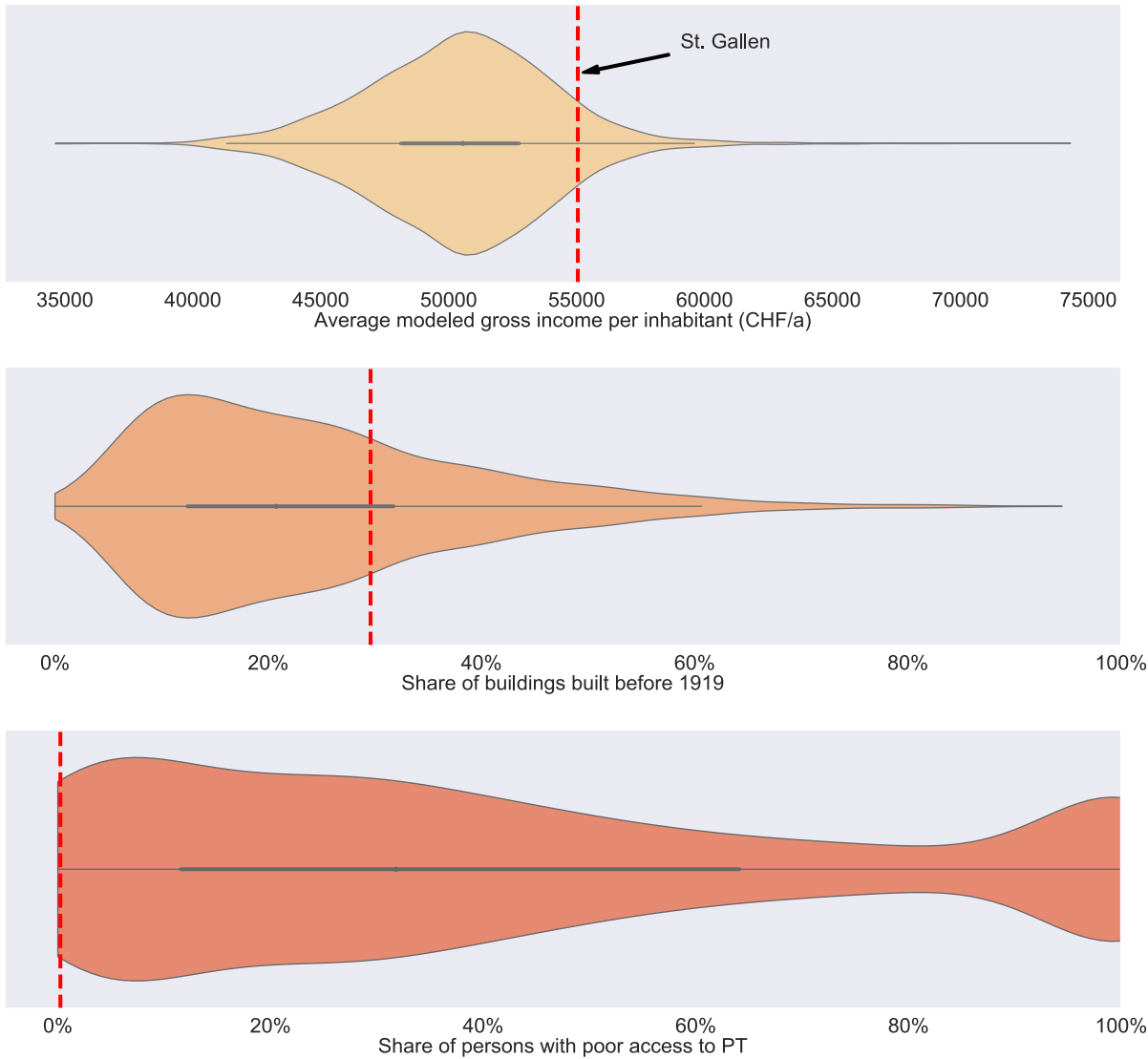

## 4.6.  Local/regional boundaries

Up to this point, we have presented results for the global planetary boundaries. In this section, we cover the results for local or regional scale boundaries. Local-scale boundaries are relevant only for specific indicators. For instance, there is no local scale planetary boundary for $CO_2$ emissions, because the effect of emitting 1t of $CO_2$ is the same regardless of the location. In contrast, consuming $1m^3$ of water will have a different impact depending on the water availability in the region.

### 4.6.1.  Freshwater use at the watershed level

Our approach in evaluating local scale safe operating space for freshwater use involves four steps: (i) quantification Swiss freshwater use at the watershed level, (ii) defining safe operating space for each watershed; (iii) deriving safe operating space for a country based on a chosen effort sharing approach (e.g., GF, EPC, etc.); (iv) evaluating whether water use from (i) is within the safe operating space defined in (iii) or not.

Water availability (after environmental and human requirements have been met) in each watershed and Swiss blue water consumption at the watershed level (from a consumption-based perspective) are shown in Figure 19. Figure 19a shows watersheds with high water availability in blue and water-scarce watersheds are shown in red. Water scarce watersheds have negative values because human water consumption (HWC) and environmental water requirements (EWR) exceed available water. Effectively, this map displays safe operating spaces (SOS) at the watershed level before allocation to any specific scale. Note that for watersheds with negative values we allocate exceedance of SOS (thus, this reverses the weight of different allocation approaches, that is approaches that usually allocate a higher share of SOS (e.g., grandfathering), now allocate a higher share of negative SOS). The derivation of safe operating space at the watershed level is explained in section 2.7.3.

Figure 19b displays spatial variation in Swiss blue water footprint. Swiss consumption exerts pressures on water resources in many other regions, particularly in East Asia and North America. A high-water footprint in a specific location is not a bad thing *per se*. Production of goods using water that originates from an abundant water region or a region with sound water management does little harm. However, it is important to identify watersheds depleted due to unsustainable water use causing environmental damage.

Therefore, in the next step, we compare Swiss water consumption in each watershed (shown in Figure 19b) with the SOS assigned to Switzerland based on a specific allocation principle. The results are presented in absolute (in Figure 20) and relative (Figure 27 in Appendix) terms. Red shades indicate watersheds where water consumption is above the allocated limit (i.e., allocated boundaries have been transgressed or overshot) and blue shades indicate watersheds with water consumption below the assigned limits (i.e., allocated boundaries have not been transgressed or undershoot). Note that light blue implies closer to the boundary and "more unsafe". Also note that in Figure 27 values between 0 and -1 denote watersheds where Swiss water footprint is above the allocated budget, but the watershed itself does not exceed SOS (i.e., there is remaining water after HWC and EWR have been met). Figure 27 highlights watersheds that are above/below the assigned SOS, however, they do not show how large the transgression of SOS in absolute terms ($m^3$ transgressed). For instance, (1-10)/10 and (100-1000)/1000 yield the same result, but the latter has greater environmental impact and therefore greater reduction potential.

Swiss water consumption in most watersheds is in line with the assigned limits, but there are also several instances where Swiss water footprint exceeds safe operating space. As shown in Figure 20 these watersheds are predominantly located in Africa, Central and East Asia.

The Nile river basin in north-eastern Africa, Haihe and Yellow river basins in northeast China, Krishna watershed in India, are amongst the most notable examples. Previous water footprint assessment for Switzerland identified more or less the same hotspots facing severe water scarcity (see, e.g., Gnehm, 2012)

Overall, these results provide additional detail to the results presented in Figure 10 and demonstrate the importance of including regional/local scale dimension. Spatially explicit information may help to design more accurate and tailored policy responses targeting specific watersheds. In this context, allocation based on grandfathering has an additional relevance, since it does not only give weight to the current total water use but also its distribution (i.e., current sourcing patterns). On the other hand, the EPC gives equal water allocation to Switzerland for the Rhine or the Mississippi watershed, thereby neglecting supply chain locations and the rationale that a higher share is sourced from domestic watersheds than the share of the global population.

Figure 19 Available water and Swiss blue water footprint at the watershed level, 2016

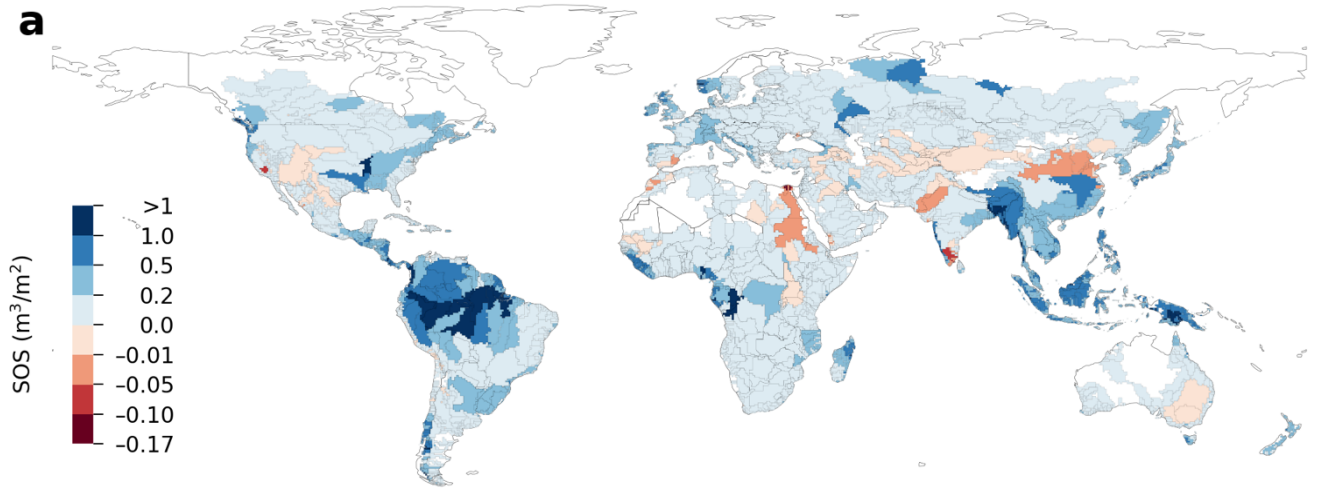

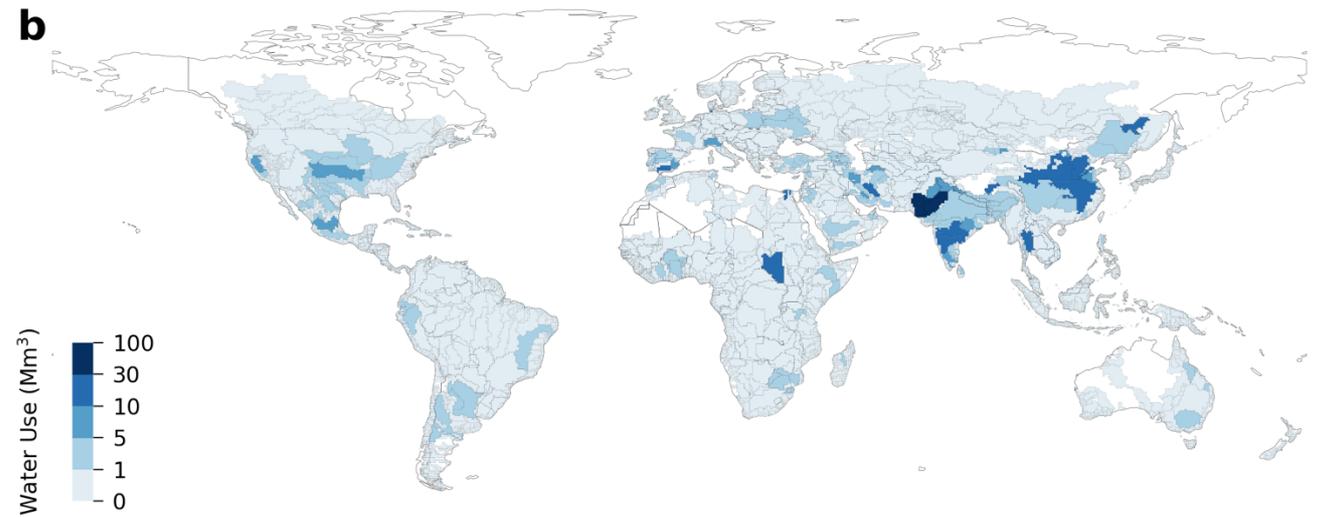

**Multiscale Orientation Values for Biodiversity, Climate and Water: A Scientific Input for Science-Based Targets**

Figure 20 Occupied safe operating space at the watershed level, in absolute terms based on three effort sharing approaches, 2016.

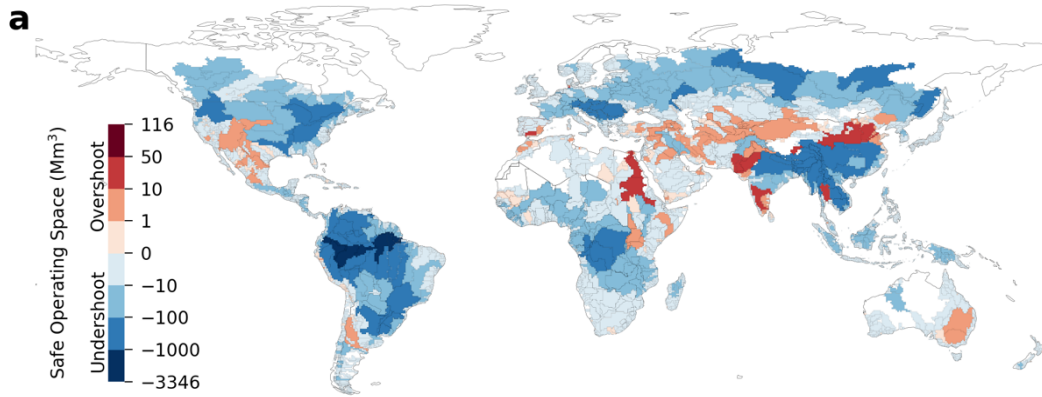

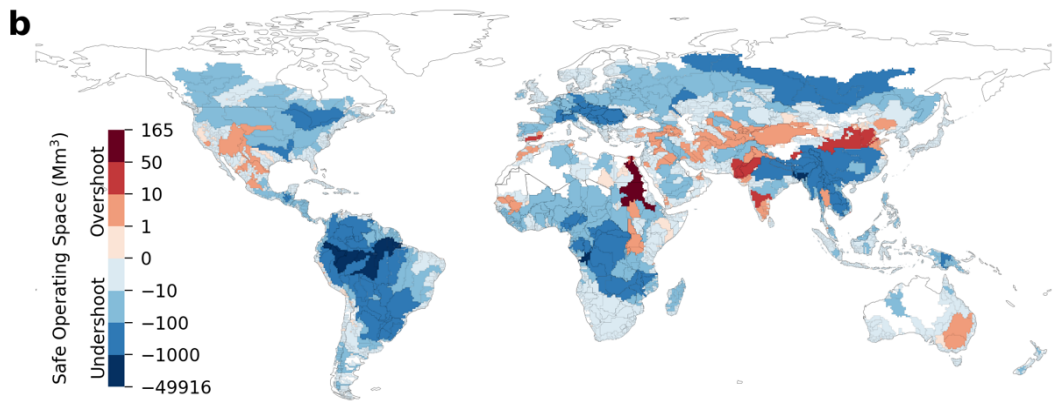

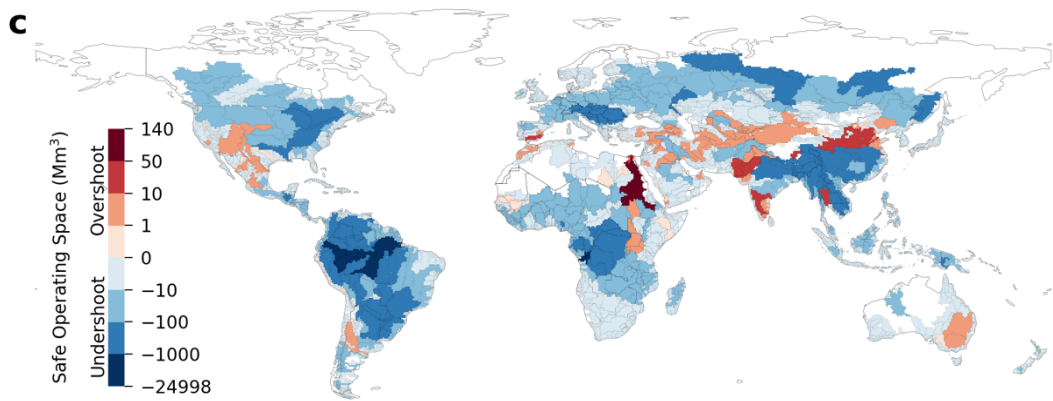

Transgression of SOS (in absolute terms) is defined as: ACTUAL-A_SOS, where A_SOS is the assigned SOS based on a specific sharing principle, ACTUAL represents water consumption in a specific watershed. Values above 0 indicate that the assigned SOS has been transgressed; values below 0 indicate that the assigned SOS has not been transgressed.

## 4.6.2. Biodiversity loss at the local level

Evaluating ecosystem functioning at the ecoregion level involves four steps: (i) quantification of Swiss land-use biodiversity loss at the ecoregion level (ii) defining safe operating space for each ecoregion, this part is based on the "nature needs half" (NNH), an aspirational goal to preserve 50% of terrestrial land areas; (iii) deriving safe operating space for a country in each ecoregion based on a chosen effort sharing approach; (iv) evaluating if the environmental impact from (i) is within the safe operating space defined in (iii).

Figure 21a shows the protection status of ecoregions of the world obtained from Dinnerstein et al. (2017). Based on these protection statuses, we derive SOS for each ecoregion, more details on this procedure are given in section 2.7.2.

Current environmental footprints are presented in Figure 21b for land-use footprint and map (c) for land use related biodiversity loss at ecoregion level. We present Figure 21b to indicate how land use translates into biodiversity loss at the ecoregion level. Dark shades indicate ecoregions with high land use and biodiversity loss footprints of Swiss consumption.

Next, we compare the assigned SOS with the actual environmental impact to evaluate the transgression of the assigned boundaries. The results for the absolute transgression of SOS are presented in Figure 22 for land use related biodiversity loss. The transgression of SOS in relative terms is shown in Figure 28. Maps in relative terms highlight ecoregions with the highest transgression levels while maps in absolute terms indicate "hotspot" ecoregions of Swiss consumption. By observing these maps, we can identify ecoregions where improvements are needed the most. Some notable examples include *West Sudanian Savanna* and *Guinean forest-savanna* ecoregion located in West Africa; *Western European Broadleaf Forests* ecoregion; and various ecoregions in Central and South Asia.

It should be emphasised that this is an experimental approach to derive and assess SOS at the ecoregion level, and the results should be interpreted with caution. In principle, we would like to have specific biodiversity targets for each ecoregion for assessing biodiversity loss at the ecoregion level. However, such targets are not available yet (or we are not aware of their existence), hence we have to rely on proxy targets. That said, protected areas are the cornerstone of biodiversity conservation, and studies document that well-managed reserves are highly effective in safeguarding biodiversity (Dinerstein et al. 2019). It should be emphasised that protected areas may involve errors related to "paper parks" (i.e., areas designated as protected but remain unprotected because of lack of enforcement).

Concerning the allocation approach, the same note for grandfathering vs EPC exists as described for freshwater use at the watershed level (previous section).

Figure 21 The protection statuses of ecoregions, Swiss land use and biodiversity footprint, 2016

The protection statuses of ecoregions of the world

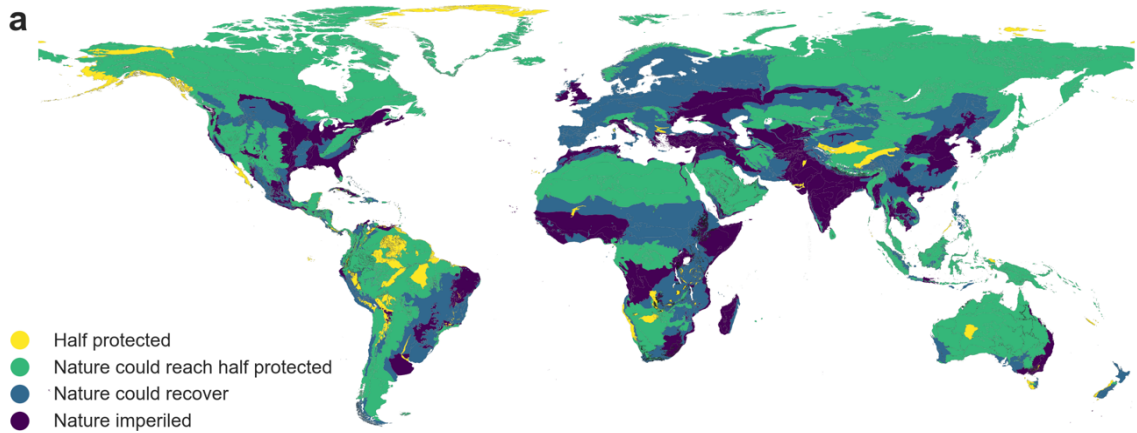

Swiss land use footprint

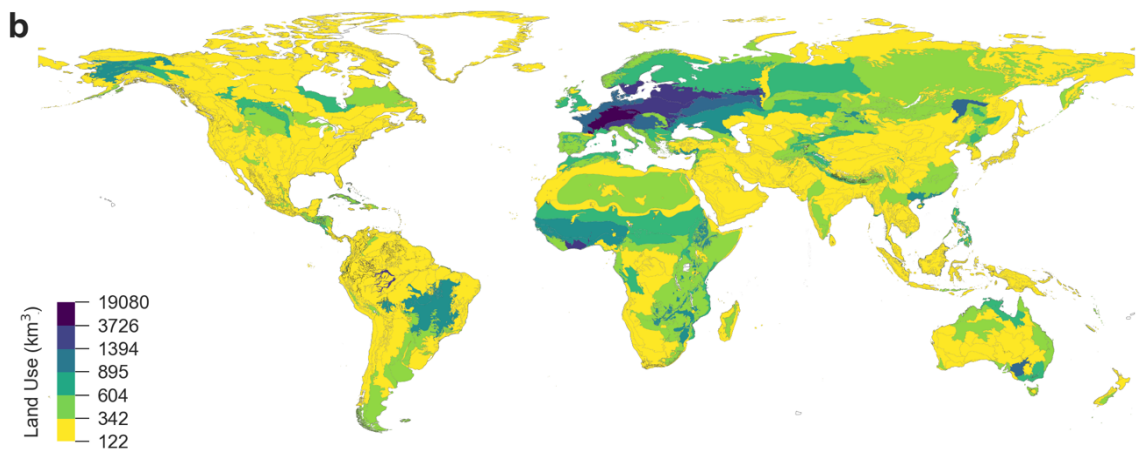

Swiss land use related biodiversity loss

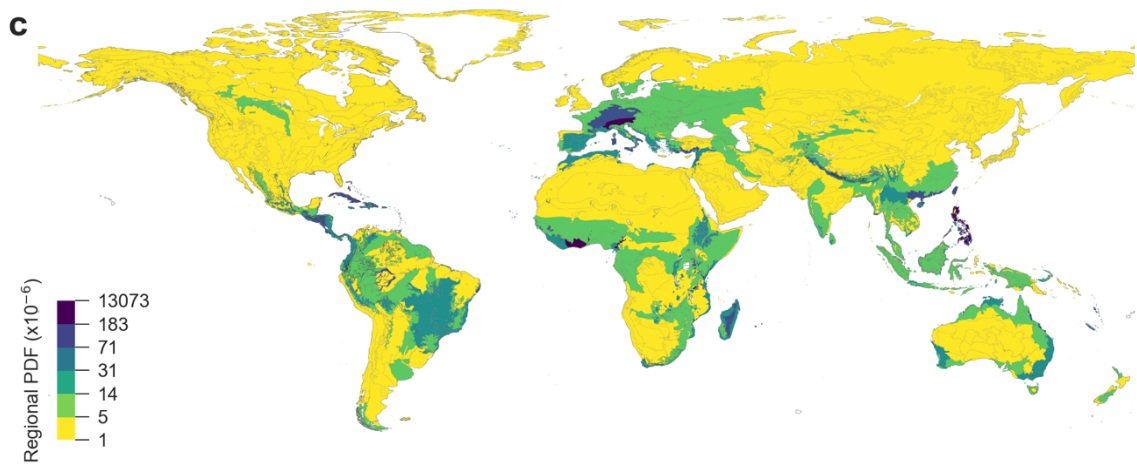

**Multiscale Orientation Values for Biodiversity, Climate and Water: A Scientific Input for Science-Based Targets**

Figure 22 Occupied safe operating space at the ecoregion level, in absolute terms based on three effort sharing approaches, 2016

Biodiversity loss: Equal per capita allocation (EPC)

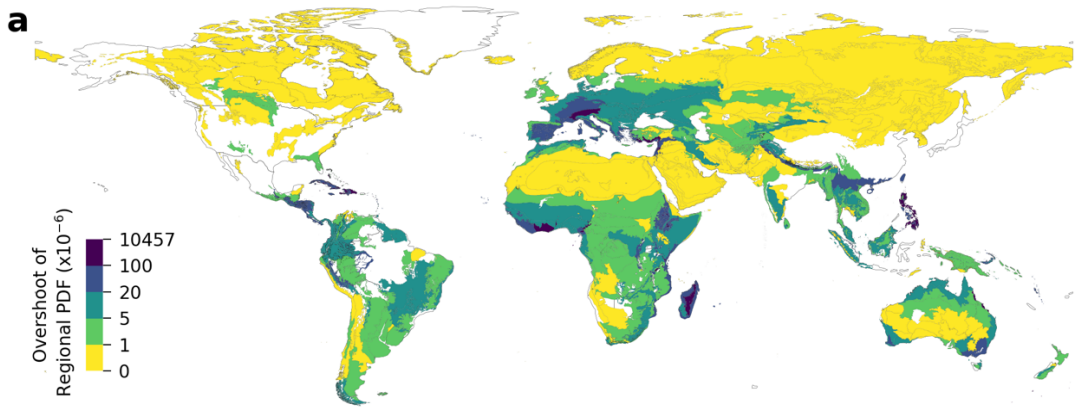

Biodiversity loss: Grandfathering allocation (GF)

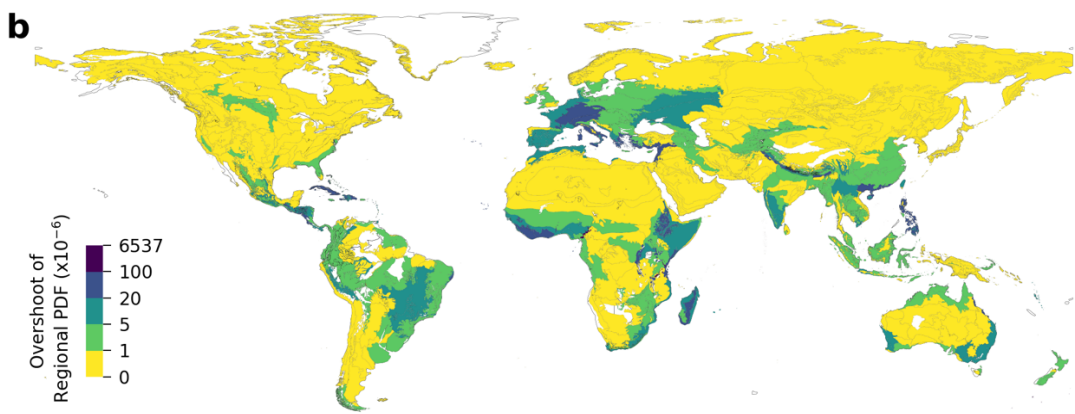

Biodiversity loss: Blended allocation (BA)

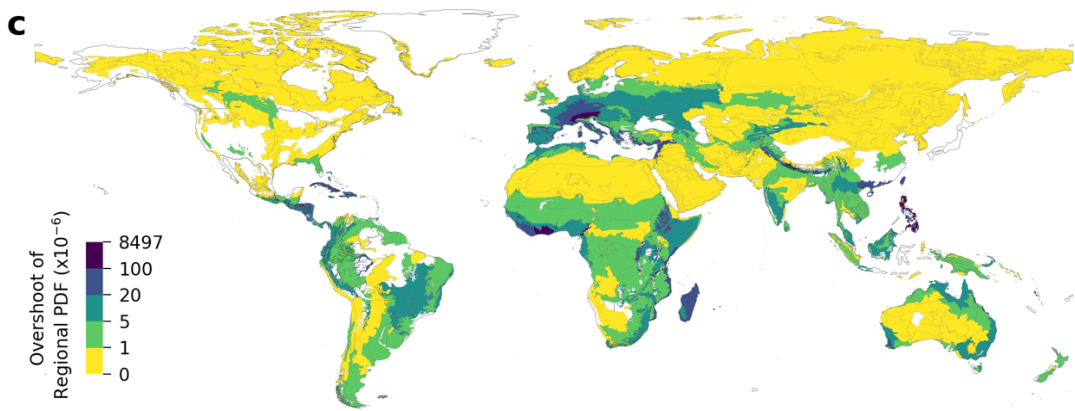

Overshoot (in absolute terms) is defined as: ACTUAL-A_SOS, where A_SOS is the assisgned SOS based on a specific sharing principle, ACTUAL represents land use related biodiversity loss in a specific ecoregion. Values above 0 indicate that the assigned SOS has been transgessed; Note that these maps only display values above 0 (i.e. only the transgessed ecoregions).

# 5. Conclusions

In this analysis, we explore a range of options and outcomes of using different allocation approaches to operationalise the PB framework to the country, sector and city scale. We demonstrate (i) how to translate the PB framework into various sub-global scales (countries, cities, industries); (ii) how to take global/local (e.g., water use at watershed level) aspects into account. Finally, to assess if Switzerland (including its industries and cities) is living within its safe operating space, we benchmarked the translated PBs for climate, biodiversity and freshwater use against actual environmental pressures from a production-based and consumption-based perspectives.

Water and biodiversity planetary boundaries do not have a well-mixed global control variable. Steffen (2015) recommend a two-tier approach to evaluate these boundaries and to account for regional-level heterogeneity. We take this into account by evaluating transgression of local and global safe operating space, where local means watershed and ecoregion levels. This approach allows to identify more specifically where the impacts occur and where the local boundaries are exceeded the most.

Identifying orientation values is challenging because various approaches for allocating fair shares to countries, sectors, and cities exist. Table 5 summarises the range of possible orientation values for the consumption perspective. Note that these orientation values apply for 2016. For climate change, given Swiss and global overshot, available budgets for reaching the 1.5°C-target are rapidly declining. If global and Swiss emissions stay the same as in 2016, the carbon budget would be exhausted within one or two decades.

Table 5 Summary of Swiss budgets for three planetary boundaries, 2016 (per capita)

| | | Consumption perspective | | | |
|---|---|---|---|---|---|
| Global Boundary | | Ability to pay | Equal per capita | Grandfathering | Blended approach |
| Climate Change (1.5°C 50%) | $tCO_2$ | 0.3 | 0.9 | 2.0 | 1.1 |
| | $tCO_2eq$ | 0.3 | 1.1 | 2.4 | 1.3 |
| Freshwater Use | $m^3$ | 150 | 534 | 804 | 496 |
| Biodiversity | Global pico PDFyr | 0.6 | 2.0 | 3.0 | 1.9 |

The results vary, and therefore we present different options but do not recommend a specific orientation value, since this requires normative choices and ethical reflections. Considering the variety of potential orientation values, the results demonstrate that Switzerland (including sectors and cities) operates beyond the "safe" limits for the climate change planetary boundary, while freshwater use is within the "safe" zone. Swiss $CO_2$, GHG emissions are above the allocated budget for all allocation approaches (considered in this study) and for both accounting methods (i.e., PBA and CBA).

In this sense, there is little ambiguity that efforts to reduce these environmental pressures need to improve. We also observe that some improvements are already taking place. For instance, $CO_2$ and total, GHG emissions, have declined since 1995. However, these declines are slow and should be enhanced. The Swiss freshwater use budget has not been overshot due to favourable climate conditions, and the trends for water consumption are decreasing. In general, freshwater use is within safe operating space. However, there are several global watersheds where Swiss water consumption contributes via imported goods to exceeding safe operating space defined at the watershed level. Furthermore, the water stress indicator, which accounts for watershed-related local scarcities, has increased since 1995 from a CBA perspective and declined from PBA perspective (the latter at a very low level).

The extent of the transgression of PBs depends on the allocation approach being used to determine safe operating space and the chosen PB limit (e.g., 1.5°C vs 2°C). Different allocation approaches yield different budgets, and the selection which approach should be used involves normative choices. Ability to Pay and Value-Added (for sectors) allocation approaches yield the lowest and highest budgets, respectively. This is expected because Switzerland (and its industries) is an outlier in terms of value-added when compared to many other countries. Other approaches such as equal per capita allocation or grandfathering are less extreme, and the blended approach yields a more balanced budget. In general, different allocation approaches will favour different type of countries/sectors, which calls for a broader discussion beyond the scientific community.

The choice of the allocation approach(es) to distribute "safe operating space" is a normative decision. Early attempts to allocate planetary boundaries have often relied on the equal per capita allocation approach, while more recent studies explored a wider range of options. EEA-FOEN (2020) used 13 allocation principles that were considered sufficient to represent the different normative choices associated with the allocation of global planetary boundaries to the European scale. Although the EEA-FOEN (2020) study has not explicitly derived any blended allocation approach, the results were benchmarked against median values, which can be considered a version of a blended approach. Blended approaches account for a wide range of principles and are likely to be more acceptable and agreeable by different actors. Also, more cooperation between scientist and policymakers is necessary to improve acceptability and, legitimacy of different allocation approaches. Finally, the global targets can only be achieved as a whole, and therefore allocation is attributing responsibility to contribute to achieving the reduction targets.

Several Planetary Boundaries are in the state of development and are likely to be updated or refined. For instance, Gleeson et al. (2020) proposed an ambitious scientific road map to define a new water planetary boundary consisting of sub-boundaries that account for various changes to the water cycle. Likewise, for the biodiversity loss, PB identifying communicable and actionable indicators and targets/limits has proven challenging. Rounsevell et al. (2020) argue that a comparable simple and measurable indicator similar to the climate target is needed to support biodiversity policy. They propose a metric based on species extinction rate with a target to keep species extinctions to well below 20 per year over the next 100 years, which is still difficult to measure or model.

Lastly, it should be acknowledged that while insights from this analysis can already be used for integrating global and local Planetary Boundaries perspectives into Swiss policy and decision-making process, due to the limitations outlined above it should be done with caution, acknowledging that these results are indicative and require refinement and improvement. Variations in studies using different data sources should not be seen as a major issue, as long as the general trend of the results is aligned. An example is the IPCC, where many climate models are used in parallel to predict an expected global warming. Even if the individual model results diverge significantly, the Paris Agreement was achieved, and such agreement could be used to set and meet, for instance biodiversity loss goals, even under uncertain indicators and thresholds.

**Multiscale Orientation Values for Biodiversity, Climate and Water: A Scientific Input for Science-Based Targets**Chaudhary A, Verones F, De Baan L, Hellweg S (2015) Quantifying Land Use Impacts on Biodiversity: Combining Species-Area Models and Vulnerability Indicators. *Environmental science & technology*, 49:9987–9995. doi: 10.1021/acs.est.5b02507

Chaudhary A, Verones F, Baan L, Pfister S, Hellweg S, (2016). Land stress: Potential species loss from land use (global; PSSRg), in: LC-IMPACT Version 1.0 A Spatially Differentiated Life Cycle Impact Assessment Approach.

https://lc-impact.eu/doc/method/Chapter11_Land_stress_July_17_2016.pdf

Crutzen, P. J. (2002). Geology of mankind. *Nature*, 415(6867), 23-23.

CONSTRAIN, (2019) ZERO IN ON the remaining carbon budget and decadal warming rates. The CONSTRAIN Project Annual Report 2019, DOI: https://doi.org/10.5518/100/20

Dao, Q. H., Peduzzi, P., Chatenoux, B., De Bono, A., Schwarzer, S., & Friot, D. (2015). Environmental limits and Swiss footprints based on planetary boundaries.

Davis, S. J., & Caldeira, K. (2010). Consumption-based accounting of CO2 emissions. *Proceedings of the National Academy of Sciences*, 107(12), 5687-5692.

Dawkins, E., André, K., Axelsson, K., Benoist, L., Swartling, Å. G., & Persson, Å. (2019). Advancing sustainable consumption at the local government level: A literature review. *Journal of Cleaner Production*, 231, 1450–1462. https://doi.org/10.1016/j.jclepro.2019.05.176

Defila R., Di Giuilo, A. (2021): Protecting Quality of Life: Protected Needs as a Point of Reference for Perceived Ethical Obligation. In: J. Martinez et al. (eds.), Handbook of Quality of Life and Sustainability, International Handbooks of Quality-of-Life, Springer Nature Switzerland, https://doi.org/10.1007/978-3-030-50540-0_13

den Elzen, M.G.J., (2002). Exploring climate regimes for differentiation of future commitments to stabilise greenhouse gas concentrations. *Integrated Assessment*, 3(4): 343-359

Dietzenbacher, E., Cazcarro, I., & Arto, I. (2020). Towards a more effective climate policy on international trade. *Nature communications*, 11(1).

Dinerstein, E., Vynne, C., Sala, E., Joshi, A. R., Fernando, S., Lovejoy, T. E., ... & Burgess, N. D. (2019). A global deal for nature: guiding principles, milestones, and targets. Science Advances, 5(4).

Dinerstein, E., Olson, D., Joshi, A., Vynne, C., Burgess, N. D., Wikramanayake, E., ... & Hansen, M. (2017). An ecoregion-based approach to protecting half the terrestrial realm. *BioScience*, 67(6), 534-545.

EEA (2013). European Union $CO_2$ emissions: different accounting perspectives. EEA Technical report No 20/2013.
Page 52 / 64

x

**Multiscale Orientation Values for Biodiversity, Climate and Water: A Scientific Input for Science-Based Targets**IRP (2019) Global Resource Outlook (GRO) report. Available at: https://www.resourcepanel.org/reports/global-resources-outlook

Jakob, M., Ward, H., & Steckel, J. C. Sharing responsibility for trade-related emissions based on economic benefits. *Global Environmental Change*, 66, 102207.

Jakob, M., & Marschinski, R. (2013). Interpreting trade-related CO2 emission transfers. *Nature Climate Change*, 3(1), 19-23.

Kander, A., Jiborn, M., Moran, D. D., & Wiedmann, T. O. (2015). National greenhouse-gas accounting for effective climate policy on international trade. *Nature Climate Change*, 5(5), 431-435.

Leontief, W. W. (1936). Quantitative input and output relations in the economic systems of the United States. *The review of economic statistics*, 105-125.

Leontief, W. (1970). Environmental repercussions and the economic structure: an input-output approach. *The review of economics and statistics*, 262-271.

Lenzen, M., Murray, J., Sack, F., & Wiedmann, T. (2007). Shared producer and consumer responsibility—Theory and practice. *Ecological economics*, 61(1), 27-42.

Lewis, S., Maslin, M. (2015) Defining the Anthropocene. *Nature* 519, 171–180. https://doi.org/10.1038/nature14258

Locke, H. (2014). Nature needs half: a necessary and hopeful new agenda for protected areas. *Nature New South Wales*, 58(3), 7.

Lucas and Wilting (2018). Using planetary boundaries to support national implementation of environment-related Sustainable Development Goals, PBL Netherlands Environmental Assessment Agency, The Hague.

Lucas, P. L., Wilting, H. C., Hof, A. F., & van Vuuren, D. P. (2020). Allocating planetary boundaries to large economies: Distributional consequences of alternative perspectives on distributive fairness. *Global Environmental Change*, 60, 102017.

Lutter S, Pfister S, Giljum S, Wieland H, Mutel C. (2016). Spatially explicit assessment of water embodied in European trade: A product-level multi-regional input-output analysis. *Global environmental change* 38: 171-182.

Marquardt, S. G., Guindon, M., Wilting, H. C., Steinmann, Z. J., Sim, S., Kulak, M., & Huijbregts, M. A. (2019). Consumption-based biodiversity footprints–Do different indicators yield different results?. *Ecological Indicators*, 103, 461-470.

Miller, R. E., & Blair, P. D. (2009). *Input-output analysis: foundations and extensions*. Cambridge university press.
Page 55 / 64

# Supplementary information

Figure 23 Boxplot components

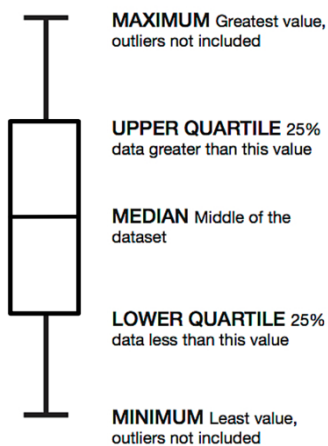

Figure 24 Production, Consumption and Territorial GHG emissions, 1995–2016

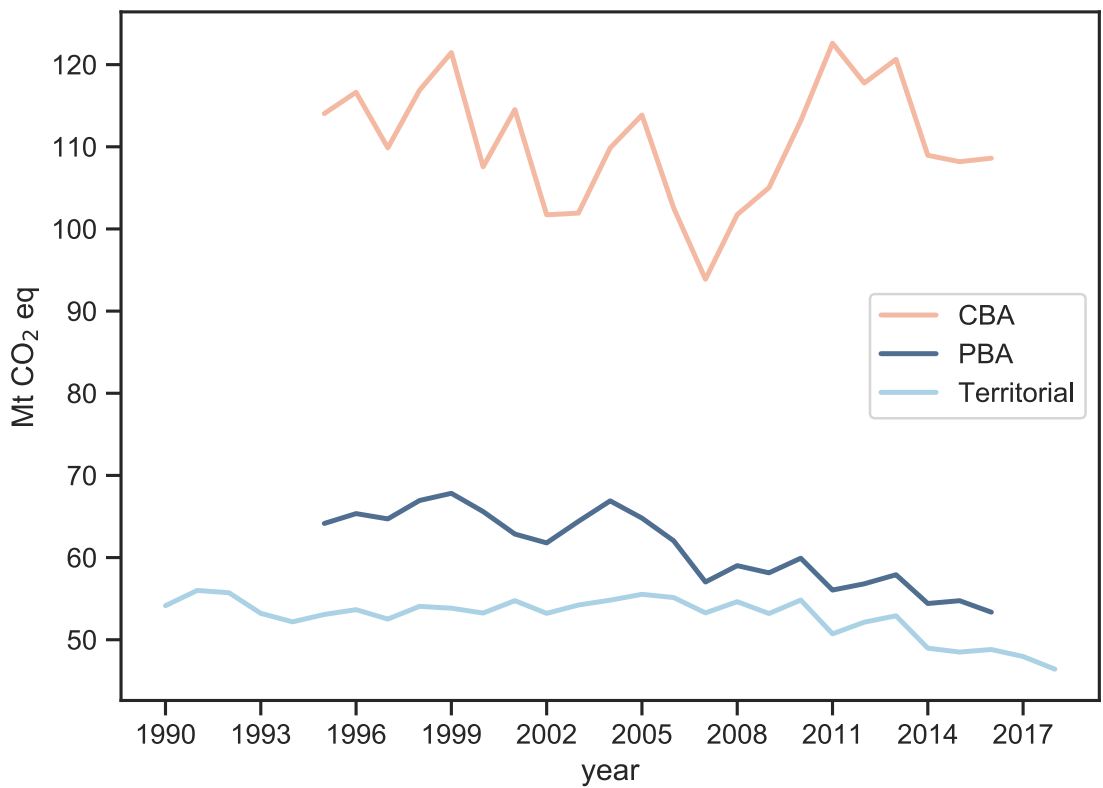

Figure 25 Swiss $CO_2$ budget for different climate targets based on five effort sharing approaches, 2016

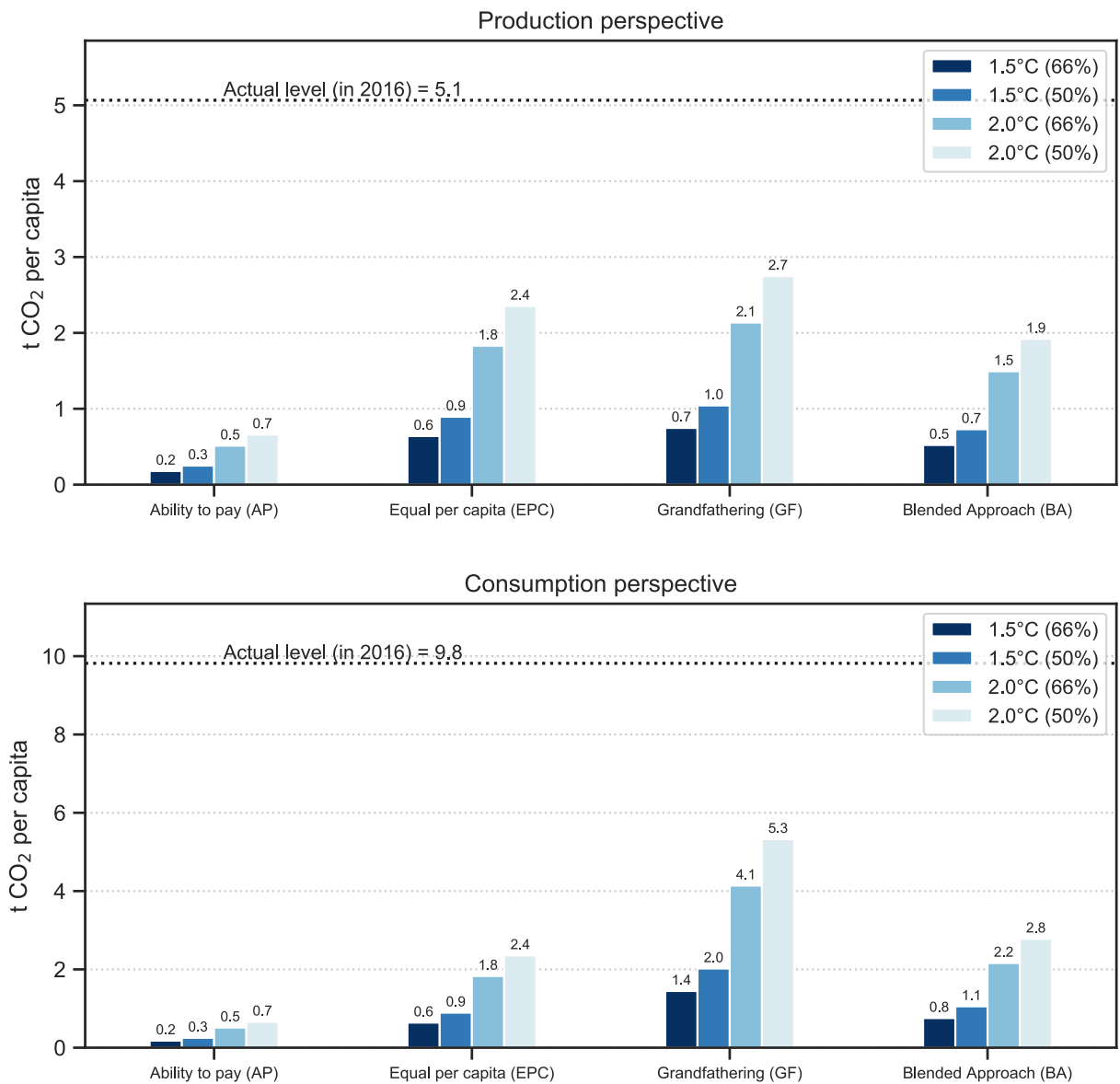

**Multiscale Orientation Values for Biodiversity, Climate and Water: A Scientific Input for Science-Based Targets**

Figure 26 Swiss GHG budget for different climate targets based on five effort sharing approaches, 2016

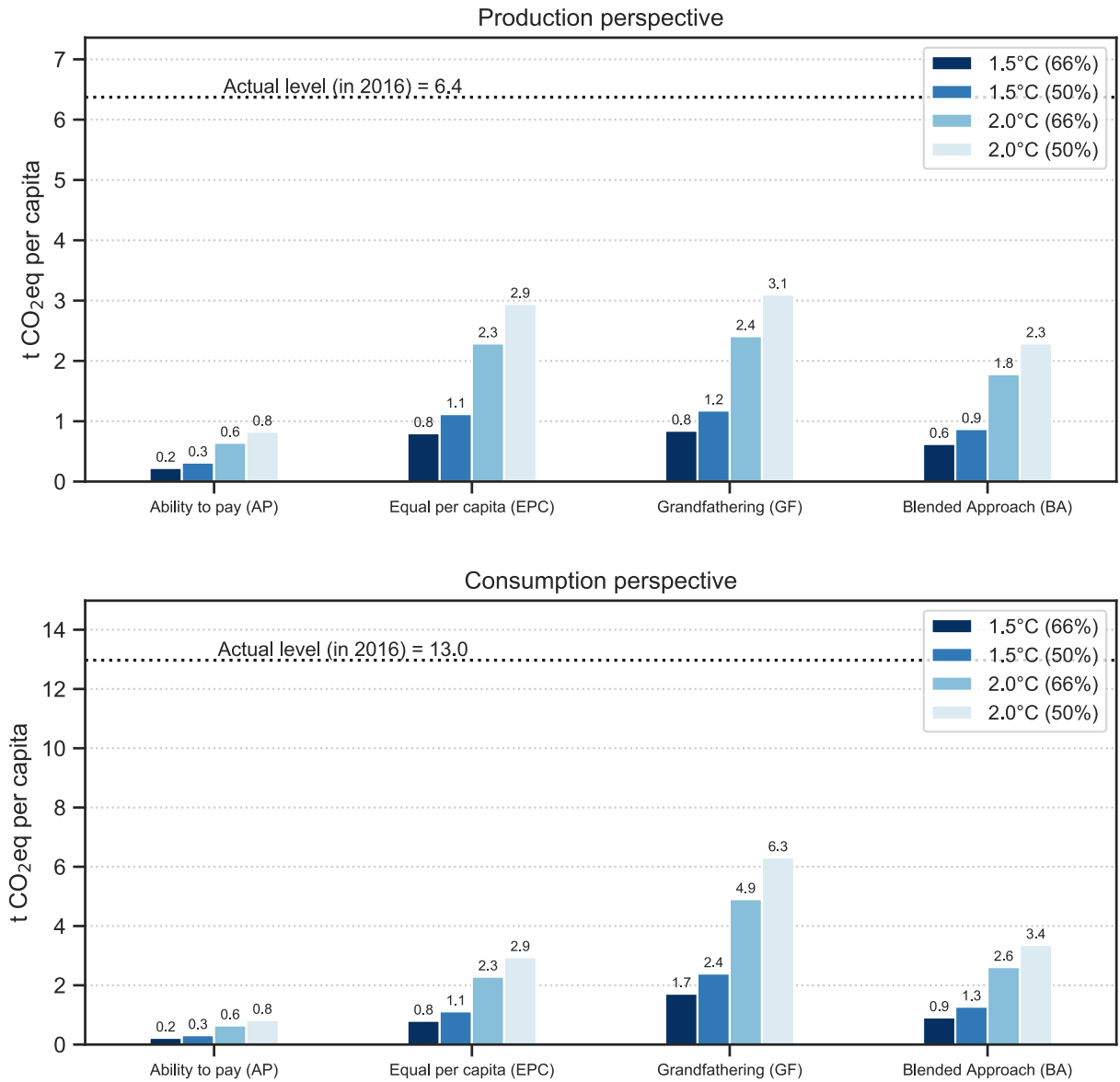

**Multiscale Orientation Values for Biodiversity, Climate and Water: A Scientific Input for Science-Based Targets**

Figure 27 Occupied safe operating space at the watershed level, in relative terms based on three effort sharing approaches, 2016

Equal per capita allocation (EPC)

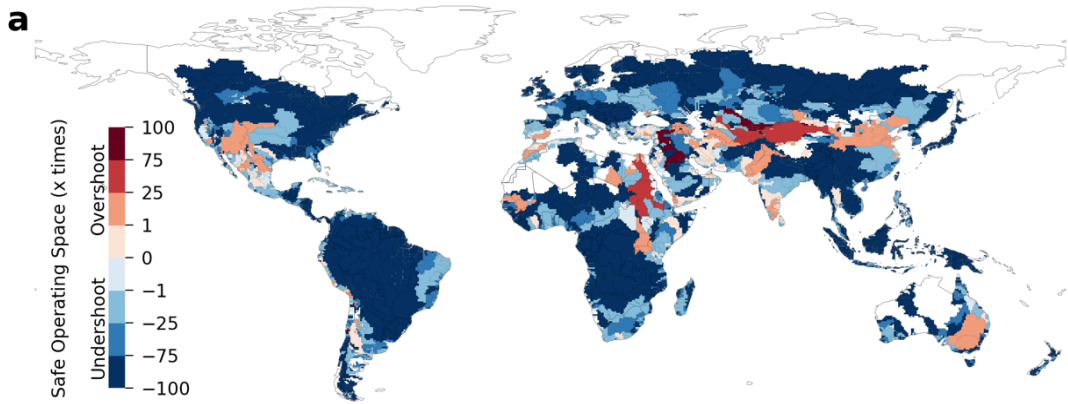

Grandfathering allocation (GF)

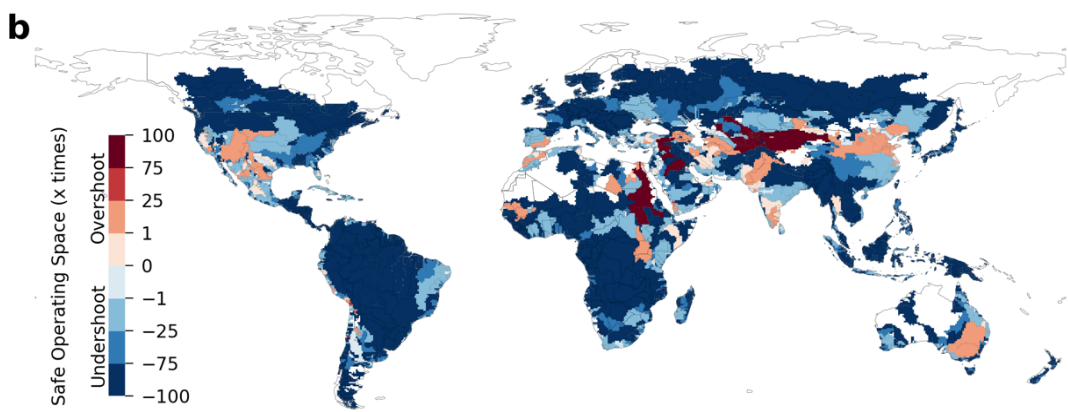

Blended allocation (BA)

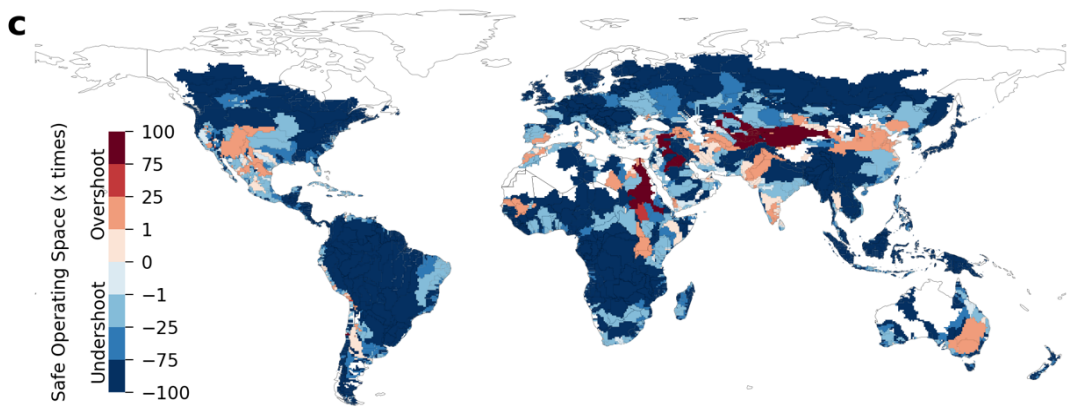

Transgression of SOS (in relative terms) is defined as: ((A_SOS-ACTUAL) / ACTUAL)*-1, where A_SOS is the assigned SOS based on a specific sharingprinciple, ACTUAL represents water consumption within a particular watershed. Values below 0 indicate that the assigned SOS has not been transgressed; values between 0-1 indicate that the assigned SOS has been transgressed, but the overall water consumption in the watershed is within SOS; values above 1 indicate that the assigned SOS has been transgressed and the overall water consumption in the watershed is above SOS. Dark blue indicates watersheds with water consumption close to the transgression of SOS.

**Multiscale Orientation Values for Biodiversity, Climate and Water: A Scientific Input for Science-Based Targets**

Figure 28 Occupied safe operating space at the ecoregion level, in relative terms based on three effort sharing approaches, 2016

Biodiversity loss: Equal per capita allocation (EPC)

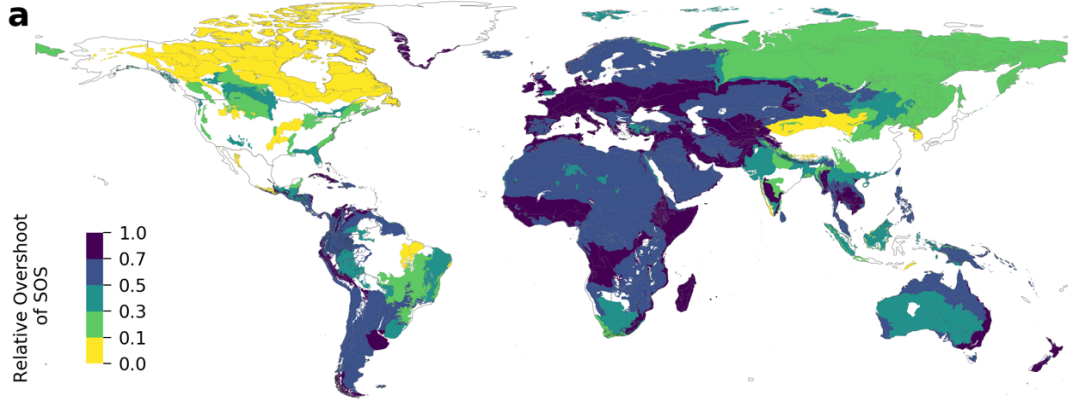

Biodiversity loss: Grandfathering allocation (GF)

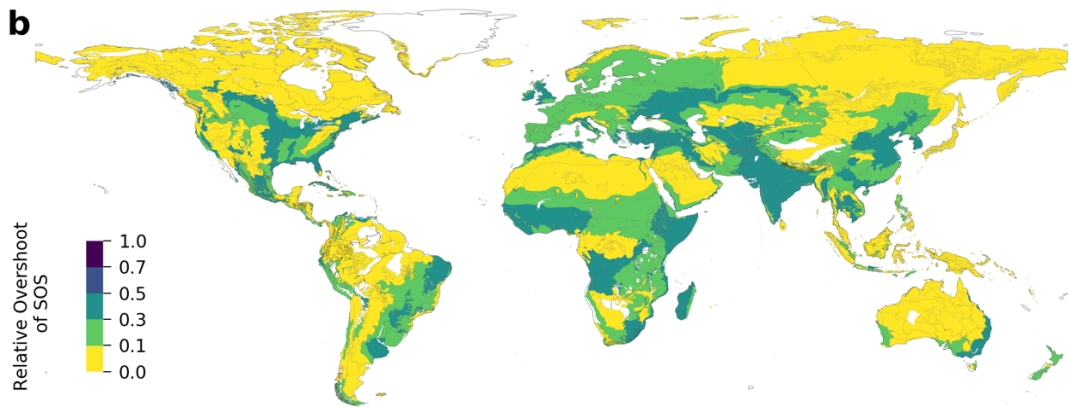

Biodiversity loss: Blended allocation (BA)

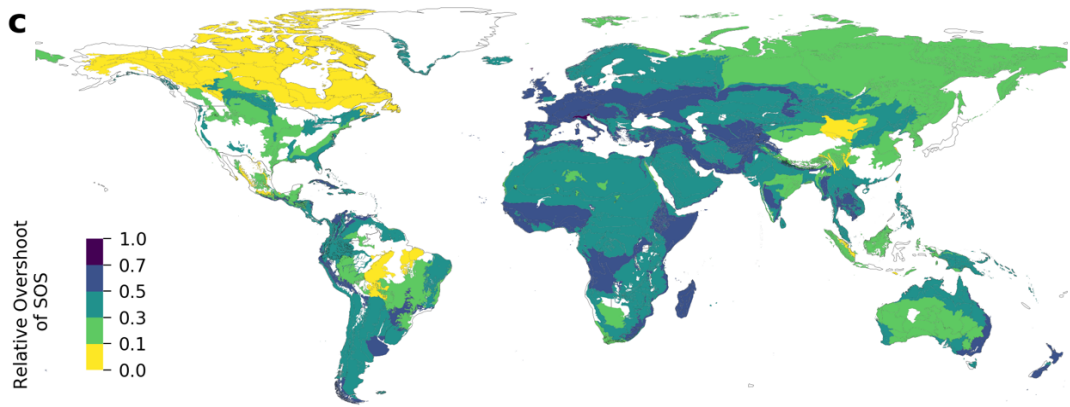

Relative Overshoot is defined as: ((A_SOS-ACTUAL) / ACTUAL)*-1, where A_SOS is the assisgned SOS based on a specific sharing principle, ACTUAL represents land use related biodiversity loss in a specific ecoregion. Values above 0 indicate that the assigned SOS has been transgessed and how much they need to be reduced to reach SOS, e.g. 0.5 implies that current impacts should be reduced by 50%; Note that these maps only display values above 0.